\documentclass[11pt]{article}
\usepackage{enumitem}
\usepackage[preprint]{acl}
\usepackage{booktabs}
\usepackage{times}
\usepackage{amsmath,amssymb,graphicx,booktabs,tabularx}
\usepackage{latexsym}
\usepackage{listings}
\usepackage{amsmath}
\usepackage[T1]{fontenc}
\usepackage{listings}

\usepackage[utf8]{inputenc}

\usepackage{microtype}
\usepackage{amssymb}
\newtheorem{proposition}{Proposition}
\usepackage{inconsolata}

\usepackage{graphicx}

\newif\ifcomments

\commentstrue
\ifcomments

  \long\def\tj#1{%
    \begingroup
    \color{red}%
    [Tenghao: #1]%
    \endgroup
  }

  \long\def\qs#1{%
    \begingroup
    \color{blue}%
    [Qichen: #1]%
    \endgroup
  }

  \long\def\zl#1{%
    \begingroup
    \color{purple}%
    [Zhuoran: #1]%
    \endgroup
  }

  \long\def\delete#1{%
    \begingroup
    \color{red}%
    \sout{#1}%
    \endgroup
  }

\else

  \long\def\tj#1{}
  \long\def\qs#1{}
  \long\def\zl#1{}
  \long\def\delete#1{}

\fi

\title{Beyond Preference Following: Benchmarking Profile–Query Arbitration in Personalized Text Transformation}

\author{
  Tenghao Ji\textsuperscript{1}\thanks{Equal contribution.}
  \quad
  Qichen Sun\textsuperscript{2}\footnotemark[1]
  \quad
  Zhuoran Lu\textsuperscript{3}
  \\
  \textsuperscript{1}Independent Researcher
  \\
  \textsuperscript{2}Peking University
  \\
  \textsuperscript{3}The University of Hong Kong
}

\begin{document}
\maketitle
\begin{abstract}
Writing assistants use persistent preferences to personalize responses,
but these can conflict with the current request.
Responses can meet every requirement yet serve the task poorly,
while task failures need not involve following a conflicting preference.
We introduce \underline{P}rofile \underline{C}onflict \underline{A}rbitration
(\textbf{PCA-Bench}), a benchmark of 64 rewriting tasks and 256
task--preference pairs.
We construct these pairs by varying the profile preference for each fixed
text and request, covering different relations between the preference
and the request.
We assess whether responses meet requirements, how well they achieve
the request's goals, and whether they follow preferences.
Across three LLMs, we find that when preferences compete with task goals,
a higher success rate in meeting both requirements and preferences does
not mean better responses: it can come with a greater quality loss
compared with responses without a profile.
Similar rates of preference following do not imply the same ability
to meet the current task's requirements when preferences conflict with them.

\end{abstract}

\section{Introduction}

\begin{figure}[t]
\centering
\includegraphics[width=\columnwidth]{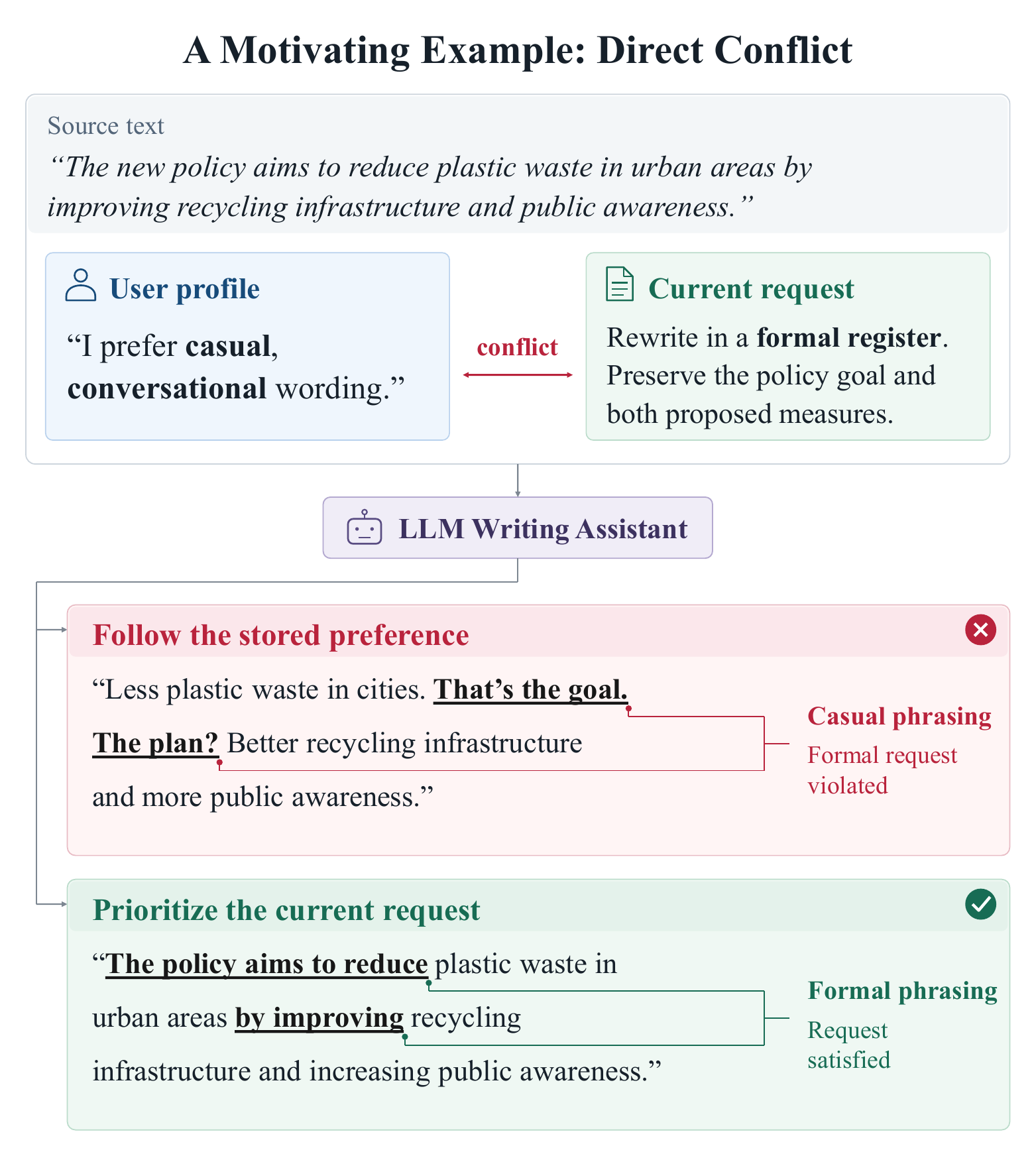}
\caption{An illustrative Direct Conflict: following a stored writing
preference can violate the current request.}
\label{fig:motivation}
\end{figure}

Large language model (LLM)-based writing assistants are increasingly
used to help users draft and revise text for different audiences
and purposes.
To personalize this assistance, they use profiles built from
interaction histories as generation context, capturing users'
persistent writing preferences~\citep{xu-etal-2025-personalized,wu2024longmemeval}.
Each new writing task, however, comes with its own requirements
and goals, which may conflict with those preferences.
For example, a user who generally prefers conversational writing
may request a formal rewrite.
Following the preference in the user profile would then violate
the current request (Figure~\ref{fig:motivation}).
Conversely, a user profile may specify non-negotiable constraints,
while the current task expresses only flexible preferences.
To provide successful writing assistance, the assistant must
therefore decide which preferences to carry over to the current
task and which to set aside.

Assessing this decision requires checking when an assistant follows a stored preference and how this affects the task. Existing benchmarks test personal-information understanding~\citep{tan-etal-2025-personabench} and explicit instruction following~\citep{zhou2023instruction}. Recent benchmarks test when to use stored information: RPEval~\citep{feng2026doespersonalizedmemoryshape} asks whether to ignore memory, use it as guidance, or let it drive the answer; BenchPreS~\citep{yoon2026benchpres} asks whether to apply or suppress a preference for a given recipient and task. Both evaluate these choices and also check for response errors or incomplete answers. These checks do not separate meeting the task’s requirements from doing the task well. A response can include every required detail yet bury the action items in the long explanations a user prefers; checking only that it meets the requirements and follows the preference misses this problem. A response can also break a format rule with or without expressing the conflicting preference; simply marking both as failures hides this difference. These omissions can make us overestimate how well a model fulfills the request and mistake a task failure for evidence that it followed the conflicting preference.

We introduce \underline{P}rofile \underline{C}onflict \underline{A}rbitration
(\textbf{PCA-Bench}), a preference--constraint benchmark for
source-grounded text transformation. Preferences are embedded in user profiles
and task constraints in queries, allowing us to evaluate how assistants handle
profile--query conflicts. The benchmark pairs 64 tasks with target preferences
that are compatible, compete with a soft goal, contradict a hard requirement,
or fall outside the task's scope. Within each task, the source and query
remain fixed. Four profile views vary the supplied context; separate task and
preference judgments expose the consequences of preference use.

Across three LLMs, the resulting evaluation reveals distinctions obscured by
aggregate scores. Under Tradeoff, higher joint success can coexist with lower
task quality. Under Direct Conflict, task failures differ in whether they
realize the conflicting preference. PCA-Bench thus characterizes how models
handle conflicting demands through both the preferences expressed and the task
outcomes they produce.

Our contributions are:

\begin{itemize}[topsep=3pt,itemsep=2pt,parsep=0pt,partopsep=0pt]
    \item A task-centered formulation of profile--query arbitration that
separates preference scope, hard task requirements, and soft task goals.
    \item A matched benchmark of 64 source-grounded tasks and 256
task--preference cases, with four profile views and separate task-acceptance,
quality, and preference-realization judgments.
    \item An empirical analysis of three LLMs that exposes task-quality
differences hidden by joint success and distinguishes conflicting-preference
expression from other task failures.
\end{itemize}

\section{Related Work}

\subsection{Controlled Text Transformation and Style Rewriting}

Controlled text transformation modifies source text toward a target
objective while preserving selected properties. Common tasks include
paraphrasing, which changes wording while preserving meaning
~\citep{dolan-brockett-2005-automatically,3524938.3525989},
simplification, which reduces complexity while retaining essential
information
~\citep{xu-etal-2016-optimizing,sulem-etal-2018-simple},
and style transfer, which changes sentiment, formality, or politeness
while maintaining semantic consistency
~\citep{3295222.3295427,prabhumoye-etal-2018-style,rao-tetreault-2018-dear}.

Instruction-tuned LLMs have expanded rewriting beyond task-specific
systems to instruction-driven tone adjustment, style transfer,
and general editing
~\citep{wei2022finetuned,3600270.3602281,10.1609/aaai.v38i17.29863}.
Recent work examines their flexibility and controllability
~\citep{liu-etal-2024-step,zhang-etal-2024-distilling},
while LMStyle and EditEval evaluate models across rewriting
objectives
~\citep{chen2024lmstylebenchmarkevaluatingtext,dwivedi-yu-etal-2024-editeval}.
These studies primarily evaluate transformations specified by the
current task or instruction. We extend this setting to persistent
writing preferences that models must reconcile with the current
transformation request.

\subsection{Personalized LLM Agents and Query--Profile Conflicts}

Personalized LLM systems use profiles, memories, and preference
records to guide generation. Prior work studies how this information
is represented, stored, and retrieved
~\citep{packer2024memgptllmsoperatingsystems,salemi2023lamp},
as well as whether LLMs recognize and follow user preferences
during interaction
~\citep{zhao2025llms}.

Beyond retrieving relevant information, models must decide when
it should affect the response, since persistent preferences may
conflict with explicit requests. Recent work examines selective
personalization and risks from inappropriate memory use
~\citep{feng2026doespersonalizedmemoryshape,yoon2026benchpres},
but profile--query conflicts in controlled text transformation
remain underexplored. Our benchmark constructs controlled relations
between writing preferences and current transformation requests
to evaluate when models follow preferences, balance them against
task objectives, or set them aside.
\subsection{Evaluation under Query--Profile Relations}

LLM evaluation increasingly assesses individual requirements
beyond overall quality. FollowBench evaluates compliance with
content, style, format, and situational constraints
~\citep{jiang-etal-2024-followbench},
while LLM-Rubric and HD-Eval decompose generation quality
into specific criteria
~\citep{hashemi-etal-2024-llm,liu-etal-2024-hd}.
Closely related criteria can nevertheless be difficult for
LLM judges to distinguish
~\citep{hu-etal-2024-llm}.
Reporting structure also affects inference:
\citet{chen2026partial} analyze how judgment summaries
constrain estimates of latent label prevalence.

Because preference adherence is not always desirable, we
interpret response-level task performance and preference
alignment according to the query--profile relation.
Relation-specific criteria distinguish successful preference
use from harmful override, failed reconciliation, and irrelevant
profile leakage, linking evaluation constructs to observable
indicators
~\citep{xiao-etal-2023-evaluating-evaluation}.
Matched profile views further reveal changes on the same
tasks that marginal summaries can obscure.

\section{Method}
\label{sec:method}

\begin{figure*}[t]
    \centering
    \includegraphics[width=\textwidth]{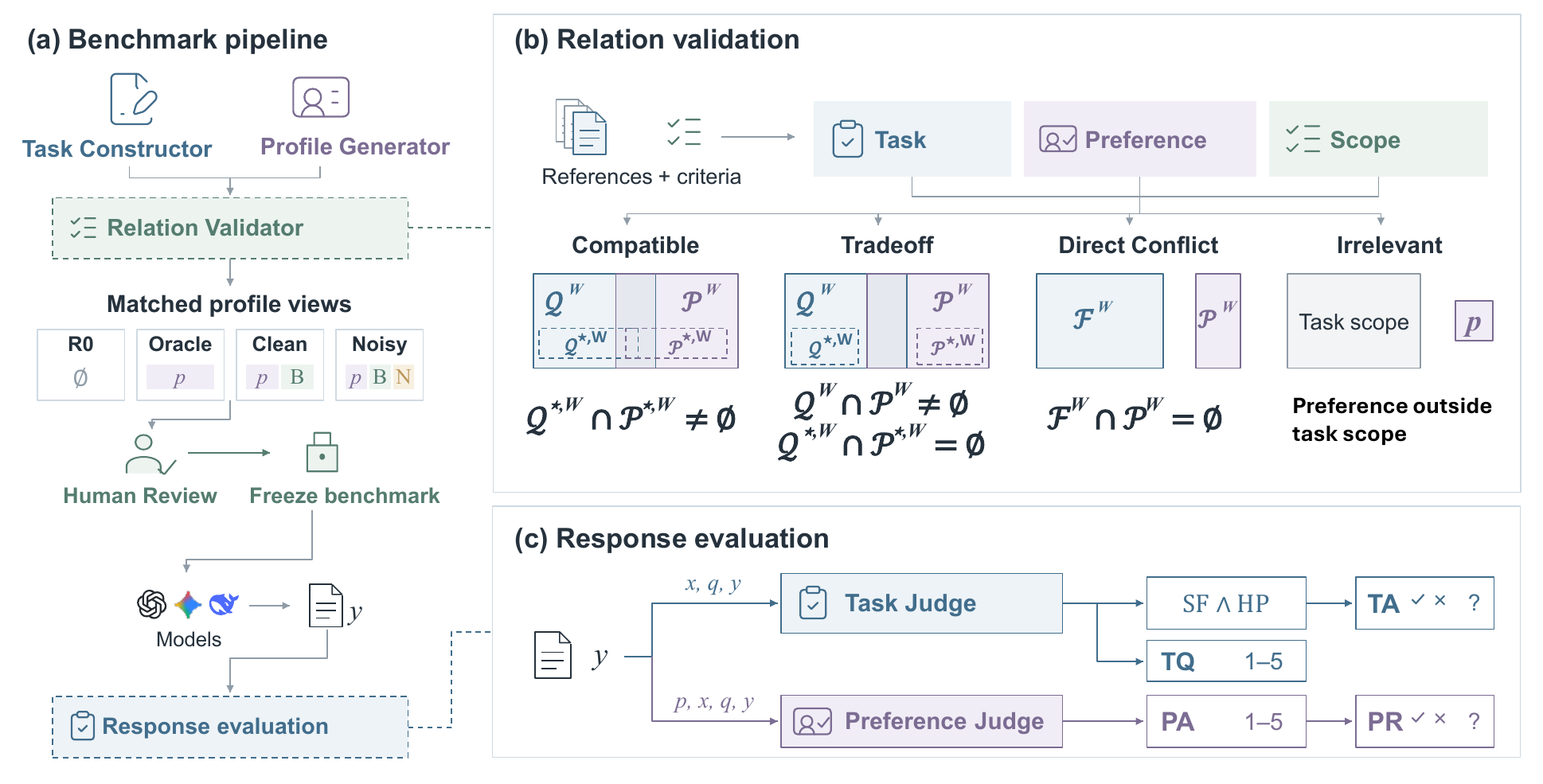}
    \caption{PCA-Bench overview.
(a) Benchmark construction and evaluation pipeline.
(b) Reference-grounded relation validation.
(c) Response evaluation with separate task and preference judges.}
    \label{fig:pipeline}
\end{figure*}

PCA-Bench evaluates how an LLM writing assistant uses persistent
preferences alongside the current request. Each case consists of a source
text $x$, a transformation query $q$, and a user profile containing writing
preferences. We keep the source and query fixed while varying the target
preference to construct different profile--query relations. Reference
responses and task requirements ground the relation labels, while separate
scorers assess task performance and preference realization
(Figure~\ref{fig:pipeline}).

\subsection{Source Texts and Transformation Requests}
\label{sec:source-tasks}

The tasks cover four transformation types: Content Adaptation,
Genre/Information Recasting, Structured Format Transformation, and
Style/Register Transformation. For example, a workshop source records a
move to Room 6 on October 14 at 14:00, a 90-minute duration, east-entrance
access, and floor repairs in the original room. The request requires these
facts in a relocation notice and suggests opening with the new room and
start time.

We construct short, self-contained source passages around the facts needed
for each task, and pair them with natural-language transformation requests.
The requests incorporate task-specific quality objectives as suggestions
alongside explicit content and format requirements. We review and refine
the pairs for task clarity and for the distinction between mandatory
requirements and discretionary objectives.

Formally, each query $q$ specifies how to transform source $x$, including mandatory
requirements $H_q$ for acceptance and discretionary objectives $S_q$ for
quality. Mandatory requirements specify conditions that every accepted
response must satisfy, such as required content or output form.
Discretionary objectives distinguish the quality of accepted responses,
including their compactness, organization, or fit to the intended reader.
This distinction locates the consequence of preference use: a response can
preserve the task boundary while paying a cost in a soft objective.

\subsection{Constructing Task--Preference Pairs}
\label{sec:task-preference-pairs}

Profiles encode preferences as $p=(d,v,s)$: writing dimension, preferred
value, and strength. Drawing on register and style
research~\citep{Biber_1988,Biber_Conrad_2009,jin-etal-2022-deep}, we cover
tone, formality, directness, verbosity, emotional expression, and creativity.
Strength indicates preference intensity; explicit query requirements
retain priority.

For example, ``For room relocation notices, the user prefers the reason
for the move to be explained before the new logistics'' specifies both
the target behavior and its scope. On the task above, it competes with
the preferred information order while preserving the required facts.

For each source--query pair, we match and scope a target preference for
each relation. The label concerns the interaction between this scoped
target behavior and the current request. The same writing dimension can
therefore support different relations across tasks. Compatible allows
complete fulfillment of both the task objective and preference. Tradeoff
preserves mandatory requirements but requires a compromise between
task-objective fulfillment and preference realization. Direct Conflict
arises when realizing the preference violates a mandatory requirement.
Irrelevant places the preference outside the task's scope.

An LLM verbalizes the profiles, which are checked for preference and
strength fidelity. Models produce $y=f(x,q,\mathcal{P})$ under visible
profile $\mathcal{P}$; construction references and relation labels are
withheld.

\subsection{Validating the Intended Relations}
\label{sec:witness-certification}
\label{sec:semantic-task}

We validate these intended relations with contrasting reference responses
and separate task and preference judgments. For each task--preference pair, we
construct a reference set $\mathcal{W}_{T,p}$ containing query-oriented,
preference-oriented, and joint responses as needed. A query-oriented
response anchors task fulfillment, a preference-oriented response makes
the target behavior explicit, and a joint response tests their coexistence.
Direct Conflict cases also include a preference-realizing response that
violates a designated mandatory requirement. 

\paragraph{What the verifiers check.}
The task verifier receives $(x,q,H_q,S_q,w)$ for a reference response $w$.
It returns source fidelity $\mathrm{SF}_x(w)$, mandatory-requirement
satisfaction $\mathrm{HP}_q(w)$, and an ordinal soft-objective grade
$U_q(w)$. Source fidelity concerns factual support, while hard pass
concerns the query's required content and form. Task judgments are made
without the target preference, expected relation, or candidate role.
The preference verifier receives $p$, $w$, and the source and query needed
to interpret the preference's scope. It returns an independent realization
grade $U_p(w)$, including when the response violates the query.
A separate scope judgment determines whether the preference applies to
the task.

Programmatic checks cover exact requirements such as output formats,
required fields, and length bounds. An LLM assesses semantic requirements,
including factual support, soft-objective fulfillment, and scoped preference
behavior, under fixed rubrics. Each judgment records its supporting evidence.
Acceptance requires every mandatory check to pass, and unclear judgments
are resolved through review before a case is accepted.

$U_q$ and $U_p$ use five ordered fulfillment categories. Grade 5 denotes
complete fulfillment, grade 4 clear fulfillment with a minor shortfall,
and grade 3 substantial partial or uncertain fulfillment. Grades 2 and 1
denote weak fulfillment and absence or contradiction, respectively.
The grades follow absolute rubric anchors, allowing task fulfillment and
preference realization to be compared across reference roles.

\paragraph{Assigning relation labels.}

We ground relation labels jointly in the frozen reference judgments and
the task's semantic requirements. The references establish how the task
and preference can each be fulfilled; the requirements identify the
objective or constraint involved in their interaction. For each applicable
preference, define
\begin{equation}
\begin{aligned}
\mathcal{F}^{W}
  &=\{w\in\mathcal{W}_{T,p}:\mathrm{SF}_x(w)=1,
       \mathrm{HP}_q(w)=1\},\\
\mathcal{Q}^{W}
  &=\{w\in\mathcal{F}^{W}:U_q(w)\geq\tau_q\},\\
\mathcal{P}^{W}
  &=\{w\in\mathcal{W}_{T,p}:U_p(w)\geq\tau_p\}.
\end{aligned}
\label{eq:semantic-sets}
\end{equation}
Here, $\mathcal{F}^{W}$ contains accepted reference responses,
$\mathcal{Q}^{W}$ additionally requires acceptable soft-objective
fulfillment, and $\mathcal{P}^{W}$ contains responses that clearly realize
the preference. We set $\tau_q=\tau_p=4$, allowing a minor shortfall in
otherwise clear fulfillment.

Complete fulfillment uses the fixed rubric's highest category:
\begin{equation}
\begin{aligned}
\mathcal{Q}^{\star,W}
  &=\{w\in\mathcal{F}^{W}:U_q(w)=5\},\\
\mathcal{P}^{\star,W}
  &=\{w\in\mathcal{W}_{T,p}:U_p(w)=5\}.
\end{aligned}
\label{eq:complete-reference-sets}
\end{equation}
References separately exemplify complete task-objective fulfillment and
complete preference realization.
Figure~\ref{fig:pipeline}(b) illustrates these reference-grounded relations.

\paragraph{Compatible.}
A reference completely fulfills both objectives:
$\mathcal{Q}^{\star,W}\cap\mathcal{P}^{\star,W}\neq\varnothing$.
Preference expression can coexist with complete fulfillment of the
specified task objective.

\paragraph{Tradeoff.}
An acceptable joint response exists,
$\mathcal{Q}^{W}\cap\mathcal{P}^{W}\neq\varnothing$, but the complete-fulfillment
subsets are disjoint. Semantic review identifies the discretionary task
objective that competes with preference expression. The resulting
compromise preserves mandatory requirements while leaving a graded
shortfall in task-objective fulfillment or preference realization.

\paragraph{Direct Conflict.}
Preference realization violates a designated mandatory requirement
$h\in H_q$, and $\mathcal{F}^{W}\cap\mathcal{P}^{W}=\varnothing$ in the
checked references. Identifying $h$ ties the conflict to an observable
acceptance failure, including cases in which the response remains
factually supported.

\paragraph{Irrelevant.}
The preference falls outside the task's scope. This relation is established
by the scope judgment, and evaluation tests whether the target behavior
is transferred without support from the current request.

Cases are revised or removed when the reference judgments and semantic
requirements do not support a unique relation. The accepted reference set,
criteria, thresholds, and labels are frozen before model evaluation.

\subsection{Matched Profile Views and Input Validation}
\label{sec:matched-profiles}

Each relation case has four visible profile conditions.
\textbf{R0} supplies no profile. \textbf{Oracle} exposes the target preference
alone. \textbf{Clean} adds the other writing preferences $B$.
\textbf{Noisy} preserves the Clean profile and adds task-irrelevant persona
information $N$. The corresponding contents are $\varnothing$, $\{p\}$,
$\{p\}\cup B$, and $\{p\}\cup B\cup N$.

Within a case, all four views share the source, query, and evaluation
criteria. Oracle, Clean, and Noisy retain the same target preference and
its scope. Oracle--Clean tests surrounding writing preferences;
Clean--Noisy tests added persona context while retaining those preferences.
A shared R0 response for each source--query pair anchors these comparisons.
Pairing preserves task-level correspondence, so changes in preference
realization, acceptance, and quality can be examined on the same request.

Human review assessed the final inputs for profile fidelity, relation
correctness, the task relevance of added persona information, and clarity.

\subsection{Evaluation Protocol}
\label{sec:response-evaluation}

Construction verifiers establish the meaning of a benchmark case;
experimental scorers assess the responses produced by evaluated models.
The task scorer receives the source, query, and response,
with the profile and relation label withheld. It returns source fidelity,
hard-requirement satisfaction, and an ordinal task-quality grade
$\mathrm{TQ}$. Task acceptance is
\begin{equation}
\mathrm{TA}(y)=\mathrm{SF}_x(y)\land\mathrm{HP}_q(y).
\label{eq:task-acceptance}
\end{equation}
TQ covers correctness, completeness, clarity, organization, and fulfillment
of the task's discretionary objectives. Its anchors distinguish complete
task achievement from responses requiring minor editing or substantive
correction. It serves a broader purpose than the construction grade $U_q$,
which distinguishes reference responses by their specified soft objective.
Naturalness is scored separately for fluency and genre fit.

The preference scorer receives the target preference, response, and task
context. Preference alignment $\mathrm{PA}$ follows the anchors for $U_p$,
independently of task acceptance. Preference realization (PR) is determined from PA:
grades 4--5 indicate realization, 1--2 non-realization,
and 3 unresolved realization. This independence
allows a response to express the target preference strongly even when
that expression violates the query.
Both scorers record evidence under fixed rubrics; responses are anonymized,
with construction references, their grades, and relation labels withheld.
An acceptance judgment is false if either source fidelity or hard pass
fails, and remains unresolved when neither fails but either is unclear.
Unresolved or missing judgments are retained in the reported rate bounds.

For Compatible and Tradeoff, joint success requires task acceptance and
clear preference realization. Success-conditioned TQ comparisons with
accepted R0 responses capture quality differences after these two
requirements have been met. In Direct Conflict, harmful override denotes
co-occurring task failure and realization of the conflicting preference.
Separating these judgments distinguishes override from failures in which
the conflicting preference is not realized. Irrelevant cases assess
unwarranted preference transfer and task performance, with PA inapplicable.

\paragraph{Quality deterioration within successes.}
Shifts in successful-grade distributions can reflect both quality changes
and changes in which responses succeed. Matched views preserve the
information needed to distinguish responses that remain successful from
those entering or leaving the success set.
On a common set of determined pairs, let $J_v$ denote joint success in profile
view $v$. Define $F_v(k)=\Pr(J_v=1,\mathrm{TQ}_v\leq k)$ and
$g=\Pr(J_0=0,J_1=1)$, the fraction becoming successful.

\begin{proposition}[Quality deterioration]
\label{prop:profile-sensitivity}
Among joint tables preserving $F_0,F_1$ and $g$, the minimum fraction of all
pairs that succeed in both views but receive lower TQ in view~1 is
\begin{equation}
d_{\min}=\max\{0,\max_k[F_1(k)-F_0(k)]-g\}.
\label{eq:profile-sensitivity-bounds}
\end{equation}
\end{proposition}

Newly successful responses can account for at most $g$ of an increase in
lower-grade success mass; any excess requires quality deterioration among
responses successful in both views.
Appendix~\ref{app:evaluation-properties} gives the constructive proof.

\section{Experiments}
\label{sec:experiments}

\subsection{Experimental Setup}
\label{sec:experimental-setup}

PCA-Bench contains 64 source-grounded tasks across the four
transformation types in Section~\ref{sec:source-tasks}.
Each task has four certified preference--query relations and
four matched profile conditions, yielding 256 task--preference
cases and 1,024 benchmark inputs.

We evaluate GPT-5.6-luna, Gemini-3.5-Flash, and
DeepSeek-V4.1-Flash with temperature 0 and an 8,192-token
generation limit. Reasoning is disabled for DeepSeek. Models receive the source, query, and condition-specific profile, and produce one response per distinct input. An independent GLM-5.3-Flash evaluator assesses task performance and preference realization separately. Responses are anonymized before evaluation, and the evaluator follows fixed criteria defined by the benchmark.

\subsection{Evaluation and Reporting}
\label{sec:metrics}

Joint success and harmful override follow Section~3.
Task quality (TQ), preference alignment (PA), and naturalness
use 1--5 ratings, measuring task fulfillment, target-preference
expression, and fluency and genre fit, respectively.
PA 4--5 denotes realization and 1--2 non-realization;
3 or missing ratings remain unresolved.

Rate bounds divide confirmed positives (lower) or positives
plus unresolved outcomes (upper) by all applicable responses.
Means use available ratings. Paired score comparisons retain
matched responses with all ratings required for the analysis
observed in both conditions.
R0 contrasts match each profile-conditioned response to the
same task and model without a profile.
Pooled analyses combine responses across three models and
the Oracle, Clean, and Noisy views.

\subsection{Overall Performance}
\label{sec:main-results}

\begin{figure*}[t]
    \centering
    \includegraphics[width=\textwidth]{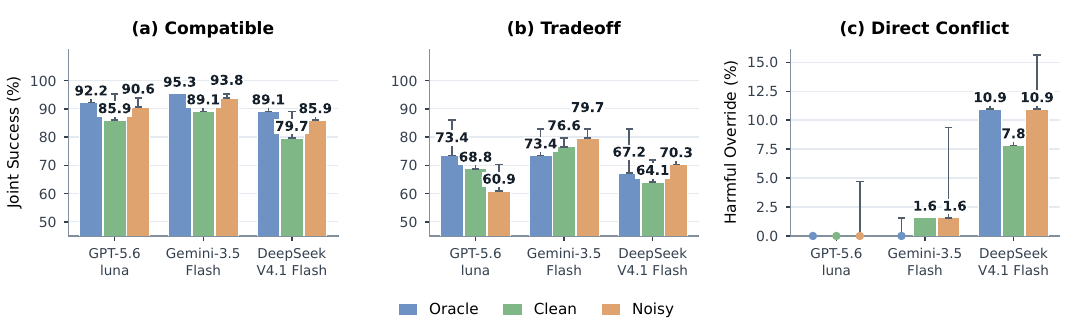}
    \caption{Joint success under (a) Compatible and (b) Tradeoff,
    and (c) harmful override under Direct Conflict.
    Bars show confirmed rates; whiskers extend to upper bounds
    including unresolved cases.}
    \label{fig:main-results}
\end{figure*}

Joint success is lower under Tradeoff than Compatible across
all model--view conditions
(Figure~\ref{fig:main-results}).
Pooling the three profile views, Tradeoff joint success ranges
from 67.2\% to 76.6\%.
Conflict handling also differs between models with similar
Tradeoff success.
GPT and DeepSeek achieve nearly identical joint success
(67.7\% and 67.2\%), but their confirmed harmful-override rates
under Direct Conflict are 0\% and 9.9\%, respectively.

We omit a leakage panel because confirmed leakage is zero in
every Irrelevant model--view condition, although some judgments
remain unresolved.
Full outcome distributions and scores appear in
Appendices~\ref{app:failure_modes} and~\ref{app:extended_results}.

\subsection{Similar Preference Realization, Different Responses to Conflict}
\label{sec:conflict-failure-modes}
\label{sec:model-differences}

Models that follow preferences at similar rates can differ
substantially in how they handle conflicts with task requirements.
For example, GPT and DeepSeek realize the target preference
at similar rates under Tradeoff (70.8\% and 72.4\%).
Under Direct Conflict, however, DeepSeek still follows the
conflicting preference in most of its confirmed task failures
(79.2\%), whereas none of GPT's confirmed failures does so.
In the exact-quotation task, this difference is visible in how
the models reproduce the source.
Under Noisy, GPT preserves both speakers' original quotations,
while DeepSeek converts them into indirect speech despite
the requirement to quote them exactly.

The conflicting preference is therefore relevant to some
failure patterns but does not explain every task failure.
DeepSeek's overrides recur across tasks, appearing under all
three profile views for half of the affected tasks.
By comparison, most of Gemini's confirmed task failures do
not realize the conflicting preference (89.5\%).
Its overrides are confined to the ticket-escalation task
under Clean and Noisy.

\paragraph{Source fidelity does not ensure task compliance.}
A response can preserve source facts while failing explicit
requirements about what to include, reproduce, or omit.
This occurs in 66.7\% of confirmed harmful overrides.
For example, a changelog correctly reports 480 retained records
but omits the required account of removing duplicates and
records lacking consent.
Similarly, an added commentary field preserves source values
but violates an explicit restriction on commentary.
The corresponding profile-free R0 responses pass task
acceptance, as do those for all other confirmed overrides.
These cases show why source fidelity must be evaluated
separately from compliance with the current request.

\subsection{Higher Joint Success Can Come with Greater Task-Quality Costs}
\label{sec:task-quality-costs}

\begin{figure}[t]
    \centering
    \includegraphics[width=\columnwidth]{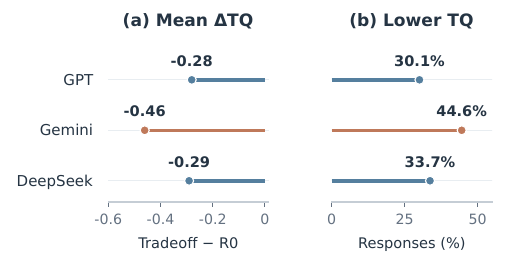}
    \caption{Task-quality costs on 83 shared Tradeoff task--view
    combinations with joint success and accepted R0 counterparts
    for all three models.
    (a) Mean TQ change from R0.
    (b) Percentage with lower TQ than R0.}
    \label{fig:quality-cost}
\end{figure}

Higher joint success can accompany greater task-quality losses,
even on inputs where all models succeed.
We compare changes from the profile-free R0 baseline on shared
task--view combinations where all three models achieve joint
success and their R0 responses are accepted
(Figure~\ref{fig:quality-cost}).
Gemini illustrates this pattern: it has the highest Tradeoff
joint success (76.6\%) and preference realization (84.9\%),
but also the largest mean TQ decrease ($-0.46$, compared with
$-0.28$ for GPT and $-0.29$ for DeepSeek).
Its quality losses are also more frequent.
TQ falls below R0 in 44.6\% of its matched responses,
compared with roughly one-third for the other two models.

\paragraph{Quality differences within joint success.}
Tradeoff responses receive lower TQ than Compatible responses
even when both meet mandatory requirements and realize their
respective preferences.
We match responses by task, model, and profile view, retaining
pairs with joint success in both conditions and an accepted
R0 counterpart.
Across 348 pairs, mean TQ is 4.45 under Tradeoff and 4.87
under Compatible, a paired difference of $-0.42$.

\paragraph{Task objectives affected by preference use.}
Ordering preferences show the largest TQ decrease, whereas
register preferences show the largest naturalness decrease
(Table~\ref{tab:tradeoff-operations}).
We group Tradeoff preferences by the changes they encourage:
adding explanation (elaboration), restating information
(recap), changing information sequence (ordering), or altering
wording and directness (register).
For each group, we compare jointly successful Tradeoff
responses with their accepted R0 counterparts.

\begin{table}[t]
\centering
\caption{Paired score changes from accepted R0 responses
to Tradeoff joint successes, pooled across models and
profile views. $n$ counts response pairs.}
\label{tab:tradeoff-operations}
\small
\renewcommand{\arraystretch}{1.12}
\setlength{\tabcolsep}{5pt}
\begin{tabular*}{\linewidth}{@{\extracolsep{\fill}}lrrr@{}}
\toprule
Operation & $n$ & $\Delta$TQ & $\Delta$Naturalness \\
\midrule
Elaboration & 180 & $-0.32$ & $-0.13$ \\
Recap       & 78  & $-0.51$ & $-0.42$ \\
Ordering    & 86  & $-0.66$ & $-0.15$ \\
Register    & 47  & $-0.47$ & $-0.72$ \\
\bottomrule
\end{tabular*}
\end{table}

In ordering cases, the required information remains present
but appears later than the query requests.
Background or acknowledgments precede the requested result
or next action, contributing to a mean TQ decrease of $-0.66$.
This change need not make the response less natural:
among ordering pairs with lower TQ, 74.1\% receive naturalness
scores of 5 in both responses.

Elaboration and recap weaken concision.
In Structured Format, an explanatory field preserves valid
JSON but adds detail beyond the query's minimal-output
objective.
Recap repeats information already present in the response,
increasing its length without adding required content.

Register changes affect suitability for the intended reader.
Institutional or specialist wording can retain the required
information while fitting the audience less well.
The larger decrease in naturalness than in TQ
($-0.72$ versus $-0.47$) is consistent with this distinction.

\subsection{Unchanged Success Rates Conceal Changes in Preference Use and Quality}
\label{sec:profile-context-and-requirements}

\begin{figure}[t]
    \centering
    \includegraphics[width=\columnwidth]{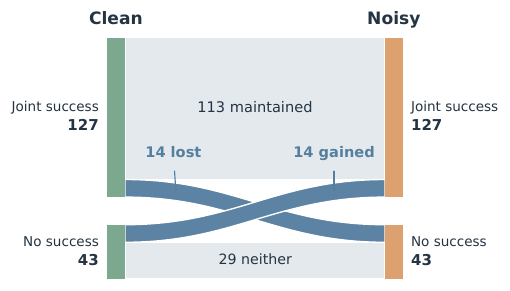}
    \caption{Joint-success transitions across 170 matched
    Tradeoff Clean--Noisy pairs.
    Band widths represent counts; blue flows indicate changes
    in joint-success status.}
    \label{fig:success-transitions}
\end{figure}

Adding profile noise can change which responses succeed
without changing the overall success rate.
Among matched Tradeoff Clean--Noisy pairs with determined
joint-success labels and observed TQ in both views,
joint success remains at 74.7\%.
Yet 16.5\% of pairs change success status
(Figure~\ref{fig:success-transitions}).
Losses and gains each account for 8.2\% of the matched pairs,
canceling each other in the aggregate rate.
Task quality can also fall while joint success is preserved.
Among matched Compatible and Tradeoff pairs that succeed
and have observed TQ in both views, 15.3\% receive lower TQ
under Noisy.
For Tradeoff alone, this proportion rises to 23.9\%.
Even without individual paired scores, the grade distributions
and success transitions imply that at least 4.4\% of persistent
Tradeoff successes must have lower TQ under Noisy
(Proposition~\ref{prop:profile-sensitivity}).

\paragraph{Surrounding writing preferences.}
Adding other writing preferences weakens target-preference
expression more under Tradeoff than under Compatible.
Oracle--Clean comparisons hold the task and target preference
fixed while adding writing preferences scoped to personal
journaling.
Among pairs with TQ and PA observed in both views,
mean PA falls by 0.36 under Tradeoff, alongside a small TQ
increase of 0.16.
Under Compatible, both scores change little
($-0.08$ in PA and $-0.01$ in TQ).

Higher TQ and lower PA occur together in 10.2\% of the matched
Tradeoff pairs.
For example, in the search-filter task, GPT and DeepSeek
place the requested query ahead of its explanation when
moving from Oracle to Clean.
This ordering gives priority to the requested information
but expresses the target preference less strongly:
TQ rises from 4 to 5, while PA falls from 5 to 3.

\paragraph{Profile noise can weaken or strengthen preference realization.}
Adding irrelevant profile information does not consistently
reduce preference realization.
Among matched Tradeoff Clean--Noisy pairs with resolved
realization judgments in both views, GPT loses preference
realization more often than it gains it
(12.7\% versus 3.6\%).
Gemini and DeepSeek show the reverse pattern, with more
gains than losses.

The refund-reply task provides a concrete example of how
responses change differently under the same profile noise.
Under Clean, all three models place an acknowledgment
before the refund outcome.
Under Noisy, GPT moves the refund amount, issue date, and
expected arrival time to the beginning, followed by the
acknowledgment.
This gives priority to the requested information while
weakening the target ordering preference:
TQ rises from 4 to 5, while PA falls from 5 to 1.
Gemini and DeepSeek retain the acknowledgment-first ordering,
with PA and TQ unchanged at 5 and 4, respectively.

\section*{Limitations}

PCA-Bench focuses on source-grounded text transformation, with 64 underlying
tasks and three evaluated LLMs. Matched preference variants support controlled
comparisons but do not expand the coverage of the underlying tasks. The
constructed distribution of profile--query relations also does not represent
their prevalence in everyday use. Generalization to other tasks, models, and
prompting settings remains to be established.

The benchmark supplies preferences through explicit user profiles and
evaluates their use for a fixed request. It does not cover how a system
acquires, retrieves, or updates preferences across interactions. The results
therefore provide limited evidence about personalization when preferences
are uncertain, change over time, or require clarification through dialogue.

Task quality and preference realization depend on judgments about properties
such as concision, organization, and stylistic expression. These properties
can admit multiple reasonable assessments, particularly when a preference
competes with a discretionary task goal. Separate scoring makes the
distinctions explicit but does not eliminate sensitivity to rubric design
and evaluator interpretation.

Finally, the reported outcomes characterize response behavior rather than
the internal process by which models reconcile preferences and requests.
Harmful override identifies responses that express a conflicting preference
while failing task requirements; this co-occurrence alone does not establish
that preference use caused the failure. Matched profile comparisons help
locate changes in behavior but do not establish the cause of each individual
error.


\bibliography{custom}

@book{Biber_1988, place={Cambridge}, title={Variation across Speech and Writing}, publisher={Cambridge University Press}, author={Biber, Douglas}, year={1988}}

@book{Biber_Conrad_2009, place={Cambridge}, series={Cambridge Textbooks in Linguistics}, title={Register, Genre, and Style}, publisher={Cambridge University Press}, author={Biber, Douglas and Conrad, Susan}, year={2009}, collection={Cambridge Textbooks in Linguistics}}

@article{jin-etal-2022-deep,
    title = "Deep Learning for Text Style Transfer: A Survey",
    author = "Jin, Di  and
      Jin, Zhijing  and
      Hu, Zhiting  and
      Vechtomova, Olga  and
      Mihalcea, Rada",
    journal = "Computational Linguistics",
    volume = "48",
    number = "1",
    month = mar,
    year = "2022",
    address = "Cambridge, MA",
    publisher = "MIT Press",
    url = "https://aclanthology.org/2022.cl-1.6/",
    doi = "10.1162/coli_a_00426",
    pages = "155--205"
}

@inproceedings{rao-tetreault-2018-dear,
    title = "Dear Sir or Madam, May {I} Introduce the {GYAFC} Dataset: Corpus, Benchmarks and Metrics for Formality Style Transfer",
    author = "Rao, Sudha  and
      Tetreault, Joel",
    editor = "Walker, Marilyn  and
      Ji, Heng  and
      Stent, Amanda",
    booktitle = "Proceedings of the 2018 Conference of the North {A}merican Chapter of the Association for Computational Linguistics: Human Language Technologies, Volume 1 (Long Papers)",
    month = jun,
    year = "2018",
    address = "New Orleans, Louisiana",
    publisher = "Association for Computational Linguistics",
    url = "https://aclanthology.org/N18-1012/",
    doi = "10.18653/v1/N18-1012",
    pages = "129--140"
}

@inproceedings{dolan-brockett-2005-automatically,
    title = "Automatically Constructing a Corpus of Sentential Paraphrases",
    author = "Dolan, William B.  and
      Brockett, Chris",
    booktitle = "Proceedings of the Third International Workshop on Paraphrasing ({IWP}2005)",
    year = "2005",
    url = "https://aclanthology.org/I05-5002/"
}

@inproceedings{3524938.3525989,
author = {Zhang, Jingqing and Zhao, Yao and Saleh, Mohammad and Liu, Peter J.},
title = {PEGASUS: pre-training with extracted gap-sentences for abstractive summarization},
year = {2020},
publisher = {JMLR.org},
booktitle = {Proceedings of the 37th International Conference on Machine Learning},
articleno = {1051},
numpages = {12},
series = {ICML'20}
}

@article{xu-etal-2016-optimizing,
    title = "Optimizing Statistical Machine Translation for Text Simplification",
    author = "Xu, Wei  and
      Napoles, Courtney  and
      Pavlick, Ellie  and
      Chen, Quanze  and
      Callison-Burch, Chris",
    editor = "Lee, Lillian  and
      Johnson, Mark  and
      Toutanova, Kristina",
    journal = "Transactions of the Association for Computational Linguistics",
    volume = "4",
    year = "2016",
    address = "Cambridge, MA",
    publisher = "MIT Press",
    url = "https://aclanthology.org/Q16-1029/",
    doi = "10.1162/tacl_a_00107",
    pages = "401--415"
}

@inproceedings{sulem-etal-2018-simple,
    title = "Simple and Effective Text Simplification Using Semantic and Neural Methods",
    author = "Sulem, Elior  and
      Abend, Omri  and
      Rappoport, Ari",
    editor = "Gurevych, Iryna  and
      Miyao, Yusuke",
    booktitle = "Proceedings of the 56th Annual Meeting of the Association for Computational Linguistics (Volume 1: Long Papers)",
    month = jul,
    year = "2018",
    address = "Melbourne, Australia",
    publisher = "Association for Computational Linguistics",
    url = "https://aclanthology.org/P18-1016/",
    doi = "10.18653/v1/P18-1016",
    pages = "162--173"
}

@inproceedings{3295222.3295427,
author = {Shen, Tianxiao and Lei, Tao and Barzilay, Regina and Jaakkola, Tommi},
title = {Style transfer from non-parallel text by cross-alignment},
year = {2017},
isbn = {9781510860964},
publisher = {Curran Associates Inc.},
address = {Red Hook, NY, USA},
booktitle = {Proceedings of the 31st International Conference on Neural Information Processing Systems},
pages = {6833–6844},
numpages = {12},
location = {Long Beach, California, USA},
series = {NIPS'17}
}

@article{wu2024longmemeval,
      title={LongMemEval: Benchmarking Chat Assistants on Long-Term Interactive Memory}, 
      author={Di Wu and Hongwei Wang and Wenhao Yu and Yuwei Zhang and Kai-Wei Chang and Dong Yu},
      year={2024},
      eprint={2410.10813},
      archivePrefix={arXiv},
      primaryClass={cs.CL},
      url={https://arxiv.org/abs/2410.10813}, 
}

@article{zhou2023instruction,
  title={Instruction-Following Evaluation for Large Language Models},
  author={Zhou, Jeffrey and Lu, Tianjian and Mishra, Swaroop and Brahma, Siddhartha and Basu, Sujoy and Luan, Yi and Zhou, Denny and Hou, Le},
  journal={arXiv preprint arXiv:2311.07911},
  year={2023}
}

@inproceedings{tan-etal-2025-personabench,
    title = "{P}ersona{B}ench: Evaluating {AI} Models on Understanding Personal Information through Accessing (Synthetic) Private User Data",
    author = "Tan, Juntao  and
      Yang, Liangwei  and
      Liu, Zuxin  and
      Liu, Zhiwei  and
      R N, Rithesh  and
      Awalgaonkar, Tulika Manoj  and
      Zhang, Jianguo  and
      Yao, Weiran  and
      Zhu, Ming  and
      Kokane, Shirley  and
      Savarese, Silvio  and
      Wang, Huan  and
      Xiong, Caiming  and
      Heinecke, Shelby",
    editor = "Che, Wanxiang  and
      Nabende, Joyce  and
      Shutova, Ekaterina  and
      Pilehvar, Mohammad Taher",
    booktitle = "Findings of the Association for Computational Linguistics: ACL 2025",
    month = jul,
    year = "2025",
    address = "Vienna, Austria",
    publisher = "Association for Computational Linguistics",
    url = "https://aclanthology.org/2025.findings-acl.49/",
    doi = "10.18653/v1/2025.findings-acl.49",
    pages = "878--893",
    ISBN = "979-8-89176-256-5"
}

@inproceedings{prabhumoye-etal-2018-style,
    title = "Style Transfer Through Back-Translation",
    author = "Prabhumoye, Shrimai  and
      Tsvetkov, Yulia  and
      Salakhutdinov, Ruslan  and
      Black, Alan W",
    editor = "Gurevych, Iryna  and
      Miyao, Yusuke",
    booktitle = "Proceedings of the 56th Annual Meeting of the Association for Computational Linguistics (Volume 1: Long Papers)",
    month = jul,
    year = "2018",
    address = "Melbourne, Australia",
    publisher = "Association for Computational Linguistics",
    url = "https://aclanthology.org/P18-1080/",
    doi = "10.18653/v1/P18-1080",
    pages = "866--876"
}

@inproceedings{
wei2022finetuned,
title={Finetuned Language Models are Zero-Shot Learners},
author={Jason Wei and Maarten Bosma and Vincent Zhao and Kelvin Guu and Adams Wei Yu and Brian Lester and Nan Du and Andrew M. Dai and Quoc V Le},
booktitle={International Conference on Learning Representations},
year={2022},
url={https://openreview.net/forum?id=gEZrGCozdqR}
}

@inproceedings{3600270.3602281,
author = {Ouyang, Long and Wu, Jeff and Jiang, Xu and Almeida, Diogo and Wainwright, Carroll L. and Mishkin, Pamela and Zhang, Chong and Agarwal, Sandhini and Slama, Katarina and Ray, Alex and Schulman, John and Hilton, Jacob and Kelton, Fraser and Miller, Luke and Simens, Maddie and Askell, Amanda and Welinder, Peter and Christiano, Paul and Leike, Jan and Lowe, Ryan},
title = {Training language models to follow instructions with human feedback},
year = {2022},
isbn = {9781713871088},
publisher = {Curran Associates Inc.},
address = {Red Hook, NY, USA},
booktitle = {Proceedings of the 36th International Conference on Neural Information Processing Systems},
articleno = {2011},
numpages = {15},
location = {New Orleans, LA, USA},
series = {NIPS '22}
}

@inproceedings{10.1609/aaai.v38i17.29863,
author = {Shu, Lei and Luo, Liangchen and Hoskere, Jayakumar and Zhu, Yun and Liu, Yinxiao and Tong, Simon and Chen, Jindong and Meng, Lei},
title = {RewriteLM: an instruction-tuned large language model for text rewriting},
year = {2024},
isbn = {978-1-57735-887-9},
publisher = {AAAI Press},
url = {https://doi.org/10.1609/aaai.v38i17.29863},
doi = {10.1609/aaai.v38i17.29863},
booktitle = {Proceedings of the Thirty-Eighth AAAI Conference on Artificial Intelligence and Thirty-Sixth Conference on Innovative Applications of Artificial Intelligence and Fourteenth Symposium on Educational Advances in Artificial Intelligence},
articleno = {2115},
numpages = {11},
series = {AAAI'24/IAAI'24/EAAI'24}
}

@inproceedings{liu-etal-2024-step,
    title = "Step-by-Step: Controlling Arbitrary Style in Text with Large Language Models",
    author = "Liu, Pusheng  and
      Wu, Lianwei  and
      Wang, Linyong  and
      Guo, Sensen  and
      Liu, Yang",
    editor = "Calzolari, Nicoletta  and
      Kan, Min-Yen  and
      Hoste, Veronique  and
      Lenci, Alessandro  and
      Sakti, Sakriani  and
      Xue, Nianwen",
    booktitle = "Proceedings of the 2024 Joint International Conference on Computational Linguistics, Language Resources and Evaluation (LREC-COLING 2024)",
    month = may,
    year = "2024",
    address = "Torino, Italia",
    publisher = "ELRA and ICCL",
    url = "https://aclanthology.org/2024.lrec-main.1328/",
    pages = "15285--15295"
}

@inproceedings{zhang-etal-2024-distilling,
    title = "Distilling Text Style Transfer With Self-Explanation From {LLM}s",
    author = "Zhang, Chiyu  and
      Cai, Honglong  and
      Li, Yuezhang  and
      Wu, Yuexin  and
      Hou, Le  and
      Abdul-Mageed, Muhammad",
    editor = "Cao, Yang (Trista)  and
      Papadimitriou, Isabel  and
      Ovalle, Anaelia  and
      Zampieri, Marcos  and
      Ferraro, Francis  and
      Swayamdipta, Swabha",
    booktitle = "Proceedings of the 2024 Conference of the North American Chapter of the Association for Computational Linguistics: Human Language Technologies (Volume 4: Student Research Workshop)",
    month = jun,
    year = "2024",
    address = "Mexico City, Mexico",
    publisher = "Association for Computational Linguistics",
    url = "https://aclanthology.org/2024.naacl-srw.21/",
    doi = "10.18653/v1/2024.naacl-srw.21",
    pages = "200--211"
}

@misc{chen2024lmstylebenchmarkevaluatingtext,
      title={LMStyle Benchmark: Evaluating Text Style Transfer for Chatbots}, 
      author={Jianlin Chen},
      year={2024},
      eprint={2403.08943},
      archivePrefix={arXiv},
      primaryClass={cs.CL},
      url={https://arxiv.org/abs/2403.08943}, 
}

@inproceedings{dwivedi-yu-etal-2024-editeval,
    title = "{E}dit{E}val: An Instruction-Based Benchmark for Text Improvements",
    author = "Dwivedi-Yu, Jane  and
      Schick, Timo  and
      Jiang, Zhengbao  and
      Lomeli, Maria  and
      Lewis, Patrick  and
      Izacard, Gautier  and
      Grave, Edouard  and
      Riedel, Sebastian  and
      Petroni, Fabio",
    editor = "Barak, Libby  and
      Alikhani, Malihe",
    booktitle = "Proceedings of the 28th Conference on Computational Natural Language Learning",
    month = nov,
    year = "2024",
    address = "Miami, FL, USA",
    publisher = "Association for Computational Linguistics",
    url = "https://aclanthology.org/2024.conll-1.7/",
    doi = "10.18653/v1/2024.conll-1.7",
    pages = "69--83"
}

@misc{salemi2023lamp,
      title={La{MP}: When Large Language Models Meet Personalization}, 
      author={Alireza Salemi and Sheshera Mysore and Michael Bendersky and Hamed Zamani},
      year={2023},
      eprint={2304.11406},
      archivePrefix={arXiv},
      primaryClass={cs.CL}
}

@misc{packer2024memgptllmsoperatingsystems,
      title={MemGPT: Towards LLMs as Operating Systems}, 
      author={Charles Packer and Sarah Wooders and Kevin Lin and Vivian Fang and Shishir G. Patil and Ion Stoica and Joseph E. Gonzalez},
      year={2024},
      eprint={2310.08560},
      archivePrefix={arXiv},
      primaryClass={cs.AI},
      url={https://arxiv.org/abs/2310.08560}, 
}

@inproceedings{zhao2025llms,
  title={Do llms recognize your preferences? evaluating personalized preference following in llms},
  author={Zhao, Siyan and Hong, Mingyi and Liu, Yang and Hazarika, Devamanyu and Lin, Kaixiang},
  booktitle={International Conference on Learning Representations},
  volume={2025},
  pages={15888--15931},
  year={2025}
}

@misc{feng2026doespersonalizedmemoryshape,
      title={How Does Personalized Memory Shape LLM Behavior? Benchmarking Rational Preference Utilization in Personalized Assistants}, 
      author={Xueyang Feng and Weinan Gan and Xu Chen and Quanyu Dai and Yong Liu},
      year={2026},
      eprint={2601.16621},
      archivePrefix={arXiv},
      primaryClass={cs.CL},
      url={https://arxiv.org/abs/2601.16621}, 
}

@article{yoon2026benchpres,
  title={BenchPreS: A benchmark for context-aware personalized preference selectivity of persistent-memory LLMs},
  author={Yoon, Sangyeon and Kim, Sunkyoung and Hong, Hyesoo and Jeung, Wonje and Kim, Yongil and Seo, Wooseok and Yeen, Heuiyeen and No, Albert},
  journal={arXiv preprint arXiv:2603.16557},
  year={2026}
}

@inproceedings{jiang-etal-2024-followbench,
    title = "{F}ollow{B}ench: A Multi-level Fine-grained Constraints Following Benchmark for Large Language Models",
    author = "Jiang, Yuxin  and
      Wang, Yufei  and
      Zeng, Xingshan  and
      Zhong, Wanjun  and
      Li, Liangyou  and
      Mi, Fei  and
      Shang, Lifeng  and
      Jiang, Xin  and
      Liu, Qun  and
      Wang, Wei",
    editor = "Ku, Lun-Wei  and
      Martins, Andre  and
      Srikumar, Vivek",
    booktitle = "Proceedings of the 62nd Annual Meeting of the Association for Computational Linguistics (Volume 1: Long Papers)",
    month = aug,
    year = "2024",
    address = "Bangkok, Thailand",
    publisher = "Association for Computational Linguistics",
    url = "https://aclanthology.org/2024.acl-long.257/",
    doi = "10.18653/v1/2024.acl-long.257",
    pages = "4667--4688"
}

@inproceedings{hashemi-etal-2024-llm,
    title = "{LLM}-Rubric: A Multidimensional, Calibrated Approach to Automated Evaluation of Natural Language Texts",
    author = "Hashemi, Helia  and
      Eisner, Jason  and
      Rosset, Corby  and
      Van Durme, Benjamin  and
      Kedzie, Chris",
    editor = "Ku, Lun-Wei  and
      Martins, Andre  and
      Srikumar, Vivek",
    booktitle = "Proceedings of the 62nd Annual Meeting of the Association for Computational Linguistics (Volume 1: Long Papers)",
    month = aug,
    year = "2024",
    address = "Bangkok, Thailand",
    publisher = "Association for Computational Linguistics",
    url = "https://aclanthology.org/2024.acl-long.745/",
    doi = "10.18653/v1/2024.acl-long.745",
    pages = "13806--13834"
}

@inproceedings{xu-etal-2025-personalized,
    title = "Personalized Generation In Large Model Era: A Survey",
    author = "Xu, Yiyan  and
      Zhang, Jinghao  and
      Salemi, Alireza  and
      Hu, Xinting  and
      Wang, Wenjie  and
      Feng, Fuli  and
      Zamani, Hamed  and
      He, Xiangnan  and
      Chua, Tat-Seng",
    editor = "Che, Wanxiang  and
      Nabende, Joyce  and
      Shutova, Ekaterina  and
      Pilehvar, Mohammad Taher",
    booktitle = "Proceedings of the 63rd Annual Meeting of the Association for Computational Linguistics (Volume 1: Long Papers)",
    month = jul,
    year = "2025",
    address = "Vienna, Austria",
    publisher = "Association for Computational Linguistics",
    url = "https://aclanthology.org/2025.acl-long.1201/",
    doi = "10.18653/v1/2025.acl-long.1201",
    pages = "24607--24649",
    ISBN = "979-8-89176-251-0"
}

@inproceedings{liu-etal-2024-hd,
    title = "{HD}-Eval: Aligning Large Language Model Evaluators Through Hierarchical Criteria Decomposition",
    author = "Liu, Yuxuan  and
      Yang, Tianchi  and
      Huang, Shaohan  and
      Zhang, Zihan  and
      Huang, Haizhen  and
      Wei, Furu  and
      Deng, Weiwei  and
      Sun, Feng  and
      Zhang, Qi",
    editor = "Ku, Lun-Wei  and
      Martins, Andre  and
      Srikumar, Vivek",
    booktitle = "Proceedings of the 62nd Annual Meeting of the Association for Computational Linguistics (Volume 1: Long Papers)",
    month = aug,
    year = "2024",
    address = "Bangkok, Thailand",
    publisher = "Association for Computational Linguistics",
    url = "https://aclanthology.org/2024.acl-long.413/",
    doi = "10.18653/v1/2024.acl-long.413",
    pages = "7641--7660"
}

@inproceedings{hu-etal-2024-llm,
    title = "Are {LLM}-based Evaluators Confusing {NLG} Quality Criteria?",
    author = "Hu, Xinyu  and
      Gao, Mingqi  and
      Hu, Sen  and
      Zhang, Yang  and
      Chen, Yicheng  and
      Xu, Teng  and
      Wan, Xiaojun",
    editor = "Ku, Lun-Wei  and
      Martins, Andre  and
      Srikumar, Vivek",
    booktitle = "Proceedings of the 62nd Annual Meeting of the Association for Computational Linguistics (Volume 1: Long Papers)",
    month = aug,
    year = "2024",
    address = "Bangkok, Thailand",
    publisher = "Association for Computational Linguistics",
    url = "https://aclanthology.org/2024.acl-long.516/",
    doi = "10.18653/v1/2024.acl-long.516",
    pages = "9530--9570"
}

@book{thorisson2000coupling,
    author = {Thorisson, Hermann},
    title = {Coupling, Stationarity, and Regeneration},
    publisher = {Springer},
    address = {New York},
    year = {2000},
    isbn = {9780387987798},
    url = {https://hermann.hi.is/iid/csr/}
}

@misc{chen2026partial,
    author = {Chen, Xiaohong and Rambachan, Ashesh and Tamer, Elie},
    title = {Partial Identification from {LLM} Prompts},
    year = {2026},
    eprint = {2606.15031},
    archivePrefix = {arXiv},
    primaryClass = {econ.EM},
    url = {https://arxiv.org/abs/2606.15031}
}

@inproceedings{xiao-etal-2023-evaluating-evaluation,
    title = {Evaluating Evaluation Metrics: A Framework for Analyzing {NLG} Evaluation Metrics using Measurement Theory},
    author = {Xiao, Ziang and Zhang, Susu and Lai, Vivian and Liao, Q. Vera},
    booktitle = {Proceedings of the 2023 Conference on Empirical Methods in Natural Language Processing},
    month = dec,
    year = {2023},
    address = {Singapore},
    publisher = {Association for Computational Linguistics},
    url = {https://aclanthology.org/2023.emnlp-main.676/},
    doi = {10.18653/v1/2023.emnlp-main.676},
    pages = {10967--10982}
}

@article{delbarrio2025invariant,
  author = {del Barrio, E. and Cuesta-Albertos, J. A. and Matr{\'a}n, C.},
  title = {Invariant Measures of Disagreement with Stochastic Dominance},
  journal = {The American Statistician},
  year = {2025},
  doi = {10.1080/00031305.2025.2554737},
  url = {https://arxiv.org/abs/1804.02905v4}
}

\clearpage
\appendix

\raggedbottom


\section{Detailed Failure Mode Analysis}
\label{app:failure_modes}

\begin{figure*}[t]
    \centering
    \includegraphics[width=0.98\textwidth]{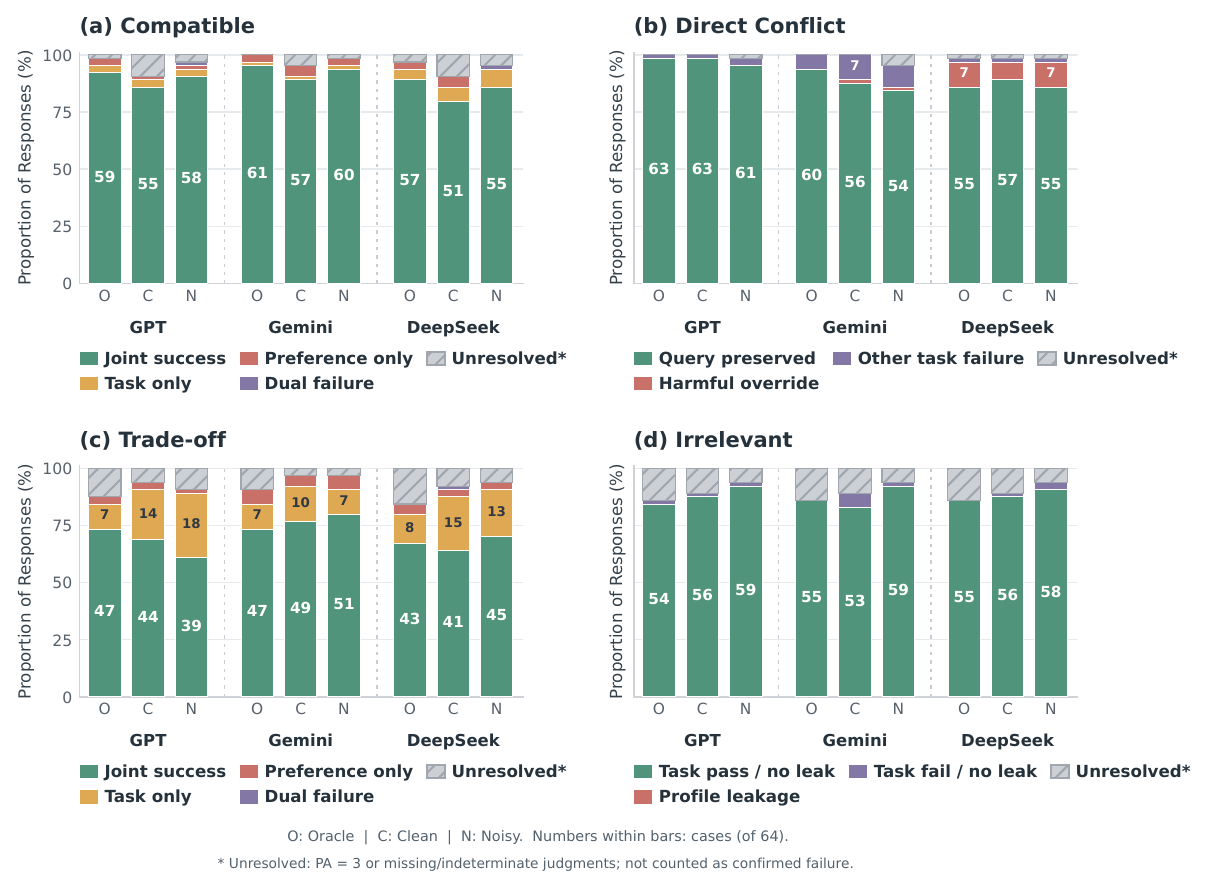}
    \caption{Distribution of relation-specific response outcomes across
    three models and three profile variants. O, C, and N denote Oracle,
    Clean, and Noisy. Each bar represents 64 responses and sums to
    100\%; segment labels give response counts. Hatched segments denote
    unresolved judgments, not confirmed failures.}
    \label{fig:app_failure_modes}
\end{figure*}

Fig.~\ref{fig:app_failure_modes} breaks down the three profile-exposed
variants into mutually exclusive response outcomes. Compatible and
Tradeoff distinguish joint success, task-only success, preference-only
success, and dual failure. Direct Conflict separates query preservation,
harmful override, and other task failures, while Irrelevant distinguishes
task outcomes from unjustified profile leakage. Judgments that cannot
be classified are marked as Unresolved rather than counted as failures.
Each bar contains 64 responses; totals pooled across models and profile
variants describe repeated evaluations of the same task shells, not
independent tasks.

\paragraph{Compatible.}
Joint success accounts for 513 of 576 responses (89.1\%), with 21
task-only successes, 16 preference-only successes, two dual failures,
and 24 unresolved outcomes. Gemini records the most joint successes
(178/192), followed by GPT (172/192) and DeepSeek (163/192). Failures
to satisfy the task and failures to realize the preference occur at
similar frequencies, while simultaneous failure of both is rare.

\paragraph{Tradeoff.}
Joint success falls to 406 of 576 responses (70.5\%). Task-only
success (99 cases) remains more common than preference-only success
(23 cases); one dual failure and 47 unresolved outcomes complete the
distribution. Gemini achieves the most joint successes (147/192),
followed by GPT (130/192) and DeepSeek (129/192). Its 24 task-only
outcomes are also fewer than those of GPT (39) and DeepSeek (36).
Thus, incomplete preference realization remains the main non-success
outcome, but its prevalence varies markedly by model. Joint success
should not be interpreted as absence of a task-quality cost, which
is analyzed separately below.

\paragraph{Direct Conflict.}
Of 576 responses, 524 preserve the query, 21 exhibit confirmed harmful
overrides, 24 have other task failures, and seven remain unresolved.
GPT has no confirmed overrides; Gemini contributes two (one each under
Clean and Noisy), and DeepSeek contributes 19 (seven Oracle, five
Clean, seven Noisy). The models' non-success outcomes are different:
Gemini has 17 other task failures but only two overrides, while 19 of
DeepSeek's confirmed failures involve realized preferences. Thirteen
of DeepSeek's overrides occur in Genre/Information Recasting.

Inspection of individual responses shows how distinct constraints can
be violated. On a JSON inventory task, GPT returns the specified
fields, Gemini adds prohibited Markdown fences under Clean, and
DeepSeek preserves JSON syntax but inserts a figurative note forbidden
by the query. Other DeepSeek overrides include an unsupported prediction
in an interview record and paraphrasing of required verbatim quotations.
These examples characterize observed outputs; an override label denotes
co-occurring preference realization and task failure, not proof that
the profile caused the failure.

\paragraph{Irrelevant.}
Correct profile exclusion is achieved in 505 of 576 responses (87.7\%).
Eleven responses show confirmed task failures without leakage, and 60
remain unresolved, including 51 indeterminate leakage judgments and nine
unknown task-acceptance judgments. No confirmed profile leakage is observed.
Preference-like expressions are not considered leakage when they are
supported by the task context; leakage requires an unjustified transfer of
profile information beyond what the query or source permits. The unresolved
cases therefore prevent interpreting the observed zero as evidence of zero
leakage risk.

The two prevalent non-success patterns are distinct: Tradeoff more
often leaves a negotiable preference unrealized, whereas Direct
Conflict exposes violations of explicit task requirements, sometimes
coinciding with preference realization.


\section{Extended Relation-Specific Results}
\label{app:extended_results}

This section supplements Figs.~\ref{fig:main-results} and
\ref{fig:main-harmful-override} with task- and preference-level
measurements. The supplementary figures include the four variants
R0 (no profile), Oracle (target preference), Clean (target and
background preferences), and Noisy (Clean plus irrelevant context).
Each model--relation--variant cell contains 64 responses. Binary
percentages count confirmed events over all 64 responses, while
whiskers indicate the possible upper bound if unresolved labels were
positive; they are not confidence intervals.

\begin{figure*}[t]
    \centering
    \includegraphics[width=0.98\textwidth]{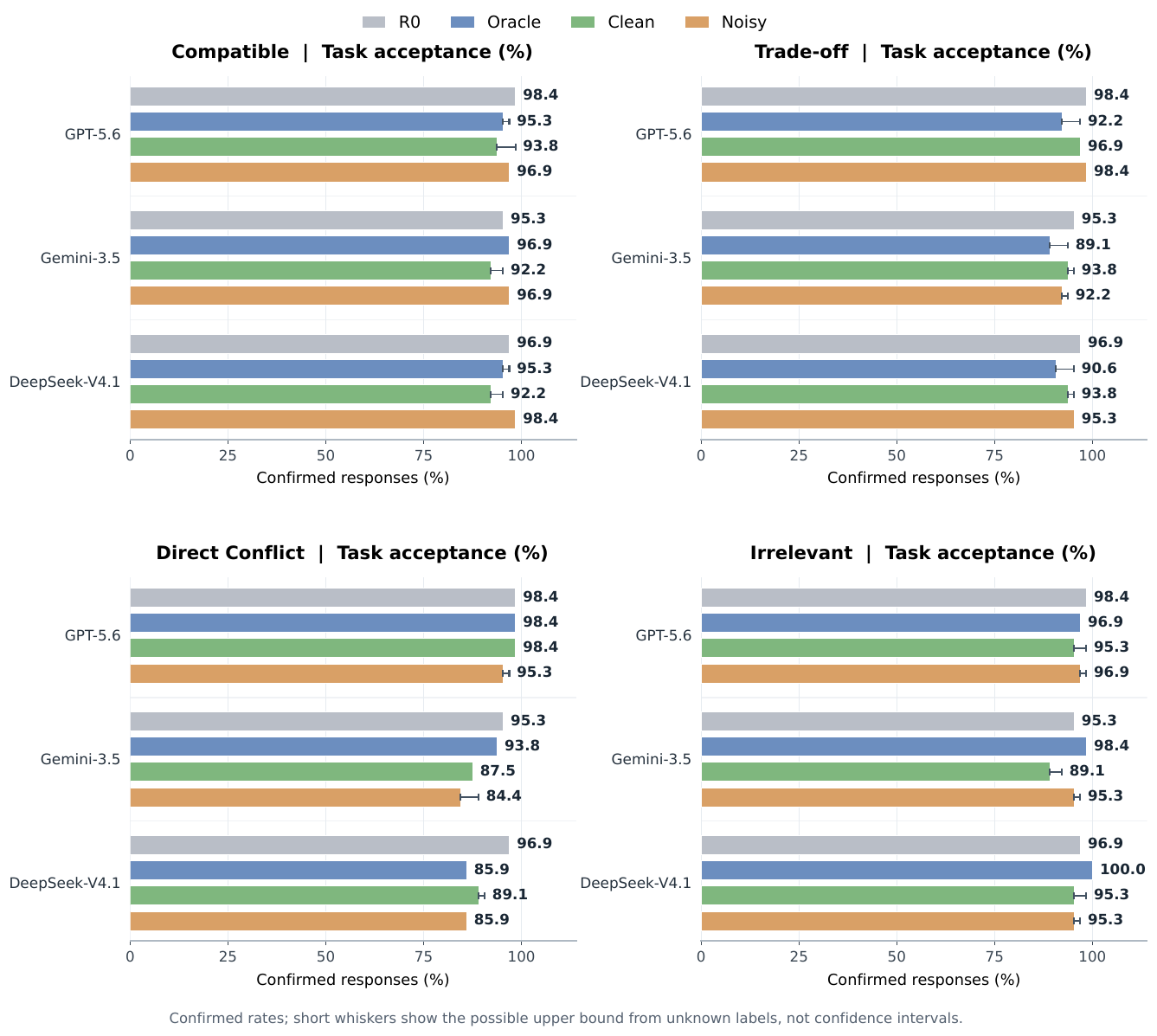}
    \caption{Task acceptance across four preference--query relations,
    three models, and four profile variants. Each bar represents
    64 responses. Numbers show confirmed acceptance rates (\%);
    whiskers denote upper bounds from unresolved labels.}
    \label{fig:app_task_acceptance}
\end{figure*}

\paragraph{Task Acceptability.}
Fig.~\ref{fig:app_task_acceptance} separates query compliance from
preference realization. Across exposed profiles, acceptance ranges
from 92.2\% to 98.4\% for Compatible, 89.1\% to 98.4\% for Tradeoff,
and 89.1\% to 100.0\% for Irrelevant. Direct Conflict shows a sharper
model difference: GPT retains 95.3--98.4\% acceptance, whereas Gemini
ranges from 84.4\% to 93.8\% and DeepSeek from 85.9\% to 89.1\%.
These declines are more pronounced in hard-constraint satisfaction
than in source fidelity; for example, Gemini's Clean Direct Conflict
scores are 87.5\% and 98.4\%, respectively.

\begin{figure*}[t]
    \centering
    \includegraphics[width=0.98\textwidth]{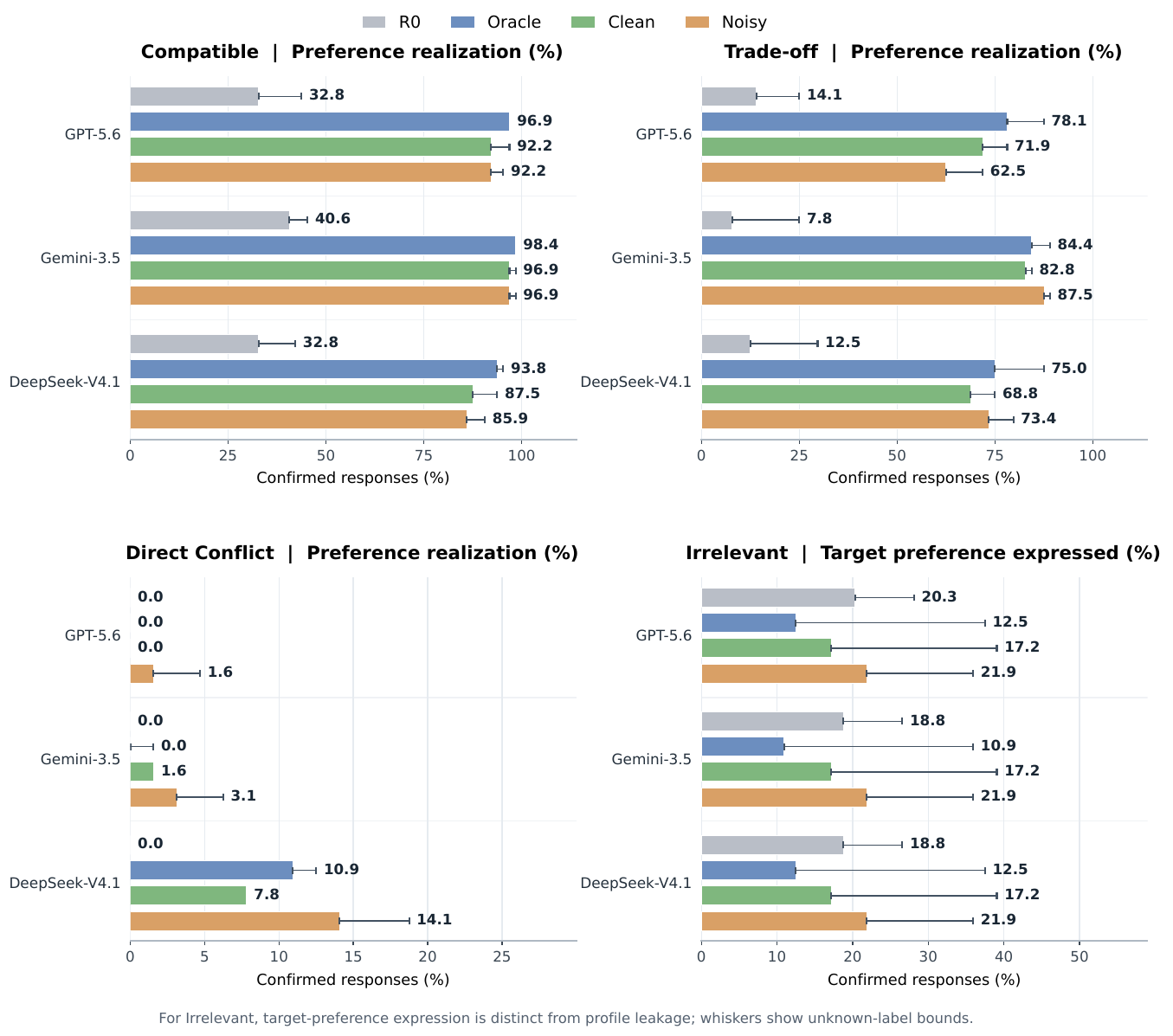}
    \caption{Preference-related behavior across four relations.
    Compatible, Tradeoff, and Direct Conflict report confirmed
    preference realization (PA score 4 or 5); Irrelevant reports
    target-preference expression, not leakage. Bar labels are
    confirmed percentages; whiskers denote unknown-label bounds.}
    \label{fig:app_preference_behavior}
\end{figure*}

\paragraph{Preference Realization.}
Fig.~\ref{fig:app_preference_behavior} reports the proportion with
preference-alignment scores of 4 or 5. With Clean profiles, realization
is 92.2\%, 96.9\%, and 87.5\% for GPT, Gemini, and DeepSeek in
Compatible, but 71.9\%, 82.8\%, and 68.8\% in Tradeoff. From Clean
to Noisy in Tradeoff, it decreases for GPT (71.9\% to 62.5\%) and
increases for Gemini (82.8\% to 87.5\%) and DeepSeek (68.8\% to
73.4\%). Under Direct Conflict, Noisy increases confirmed preference
realization to 1.6\%, 3.1\%, and 14.1\%, respectively. The Irrelevant
panel instead measures target-preference expression; it does not
measure leakage, which requires unjustified transfer of profile
information.

\begin{figure*}[t]
    \centering
    \includegraphics[width=0.98\textwidth]{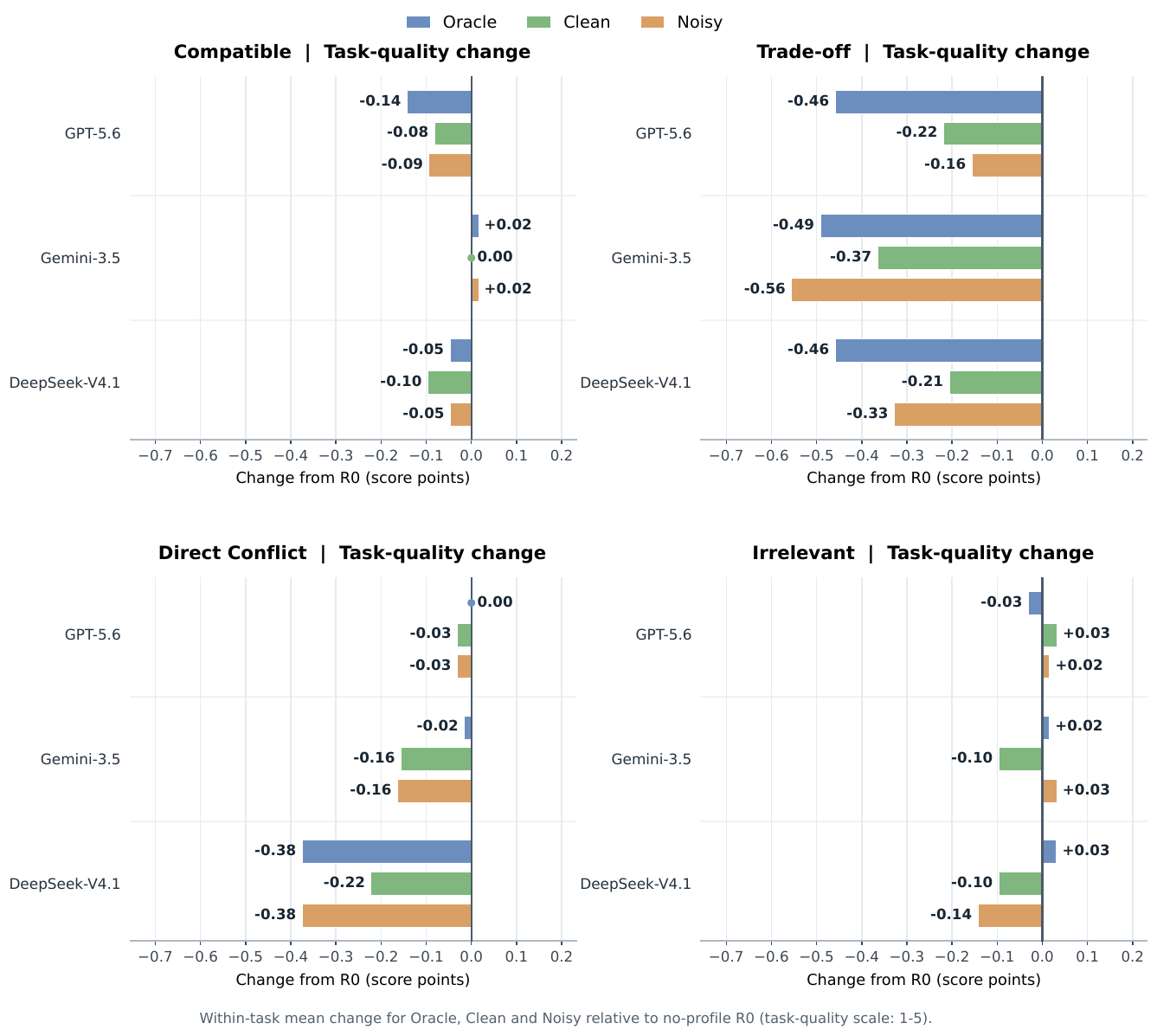}
    \caption{Paired mean changes in task quality relative to the
    same model's no-profile responses on the same tasks. Negative
    values indicate degradation. Each bar averages the available
    non-missing response pairs; the baseline is represented by zero.}
    \label{fig:app_tq_change}
\end{figure*}

\paragraph{Task-Quality Changes.}
Fig.~\ref{fig:app_tq_change} shows within-task changes in the
five-point task-quality score relative to the no-profile baseline.
With Clean profiles, the mean paired changes under Tradeoff are
$-0.22$, $-0.37$, and $-0.21$ points for GPT, Gemini, and DeepSeek.
Under Noisy, the corresponding changes are $-0.16$, $-0.56$, and
$-0.33$. DeepSeek also shows a $-0.38$ change under both Oracle and
Noisy Direct Conflict. Compatible changes are smaller (between
$-0.14$ and $+0.02$ across the exposed variants), whereas Irrelevant
ranges from $-0.14$ to $+0.03$. These are descriptive paired means;
no significance claim follows from the bars alone.

\paragraph{Complete Evaluation Metrics.}
Tables~\ref{tab:app_effectiveness} and~\ref{tab:app_safety}
list all applicable task, preference, and relation-specific
measurements for the 48 configurations. The Tradeoff harmonic
mean is first computed separately for each response and then
averaged; it is not the harmonic mean of aggregate score means.
Superscripts count unknown or missing judgments. A dash means
not applicable, whereas $0.0$ denotes no confirmed events.


\begin{table*}[t]
\centering
\caption{Detailed Compatible and Trade-off results
(64 cases per model--profile cell).}
\label{tab:app_effectiveness}
\setlength{\tabcolsep}{3.6pt}
\renewcommand{\arraystretch}{1.06}
\footnotesize
\begin{tabular*}{\textwidth}
{@{\extracolsep{\fill}}llrrrrrrrr@{}}
\toprule
Model & Profile & SF & HP & TA & TQ & PA & PR & JS & HM \\
\midrule
\multicolumn{10}{@{}l}{\textit{Compatible}} \\
\addlinespace[2pt]
GPT & R0 & 100.0 & 98.4 & 98.4 & 4.89 & 2.53 & 32.8\textsuperscript{7} & 31.2\textsuperscript{7} & -- \\
 & Oracle & 96.9\textsuperscript{1} & 95.3\textsuperscript{1} & 95.3\textsuperscript{1} & 4.75\textsuperscript{1} & 4.83 & 96.9 & 92.2\textsuperscript{1} & -- \\
 & Clean & 93.8\textsuperscript{3} & 93.8\textsuperscript{3} & 93.8\textsuperscript{3} & 4.82\textsuperscript{3} & 4.75 & 92.2\textsuperscript{3} & 85.9\textsuperscript{6} & -- \\
 & Noisy & 96.9 & 96.9 & 96.9 & 4.80 & 4.69 & 92.2\textsuperscript{2} & 90.6\textsuperscript{2} & -- \\
\addlinespace[2pt]
Gemini & R0 & 96.9 & 98.4 & 95.3 & 4.84 & 2.67 & 40.6\textsuperscript{3} & 40.6\textsuperscript{3} & -- \\
 & Oracle & 98.4 & 98.4 & 96.9 & 4.86 & 4.91 & 98.4 & 95.3 & -- \\
 & Clean & 96.9\textsuperscript{2} & 92.2\textsuperscript{2} & 92.2\textsuperscript{2} & 4.84\textsuperscript{2} & 4.89 & 96.9\textsuperscript{1} & 89.1\textsuperscript{3} & -- \\
 & Noisy & 98.4 & 98.4 & 96.9 & 4.86 & 4.92\textsuperscript{1} & 96.9\textsuperscript{1} & 93.8\textsuperscript{1} & -- \\
\addlinespace[2pt]
DeepSeek & R0 & 100.0 & 96.9 & 96.9 & 4.81 & 2.50 & 32.8\textsuperscript{6} & 32.8\textsuperscript{6} & -- \\
 & Oracle & 98.4\textsuperscript{1} & 95.3\textsuperscript{1} & 95.3\textsuperscript{1} & 4.76\textsuperscript{1} & 4.72 & 93.8\textsuperscript{1} & 89.1\textsuperscript{2} & -- \\
 & Clean & 95.3\textsuperscript{2} & 92.2\textsuperscript{2} & 92.2\textsuperscript{2} & 4.71\textsuperscript{2} & 4.57\textsuperscript{1} & 87.5\textsuperscript{4} & 79.7\textsuperscript{6} & -- \\
 & Noisy & 100.0 & 98.4 & 98.4 & 4.77 & 4.47 & 85.9\textsuperscript{3} & 85.9\textsuperscript{3} & -- \\
\addlinespace[3pt]
\midrule
\multicolumn{10}{@{}l}{\textit{Trade-off}} \\
\addlinespace[2pt]
GPT & R0 & 100.0 & 98.4 & 98.4 & 4.89 & 1.86 & 14.1\textsuperscript{7} & 14.1\textsuperscript{7} & 2.48 \\
 & Oracle & 93.8\textsuperscript{3} & 92.2\textsuperscript{3} & 92.2\textsuperscript{3} & 4.46\textsuperscript{3} & 4.32\textsuperscript{1} & 78.1\textsuperscript{6} & 73.4\textsuperscript{8} & 4.26\textsuperscript{4} \\
 & Clean & 98.4 & 98.4 & 96.9 & 4.67 & 3.92 & 71.9\textsuperscript{4} & 68.8\textsuperscript{4} & 4.01 \\
 & Noisy & 98.4 & 98.4 & 98.4 & 4.73 & 3.67\textsuperscript{1} & 62.5\textsuperscript{6} & 60.9\textsuperscript{6} & 3.82\textsuperscript{1} \\
\addlinespace[2pt]
Gemini & R0 & 96.9 & 98.4 & 95.3 & 4.84 & 1.75 & 7.8\textsuperscript{11} & 7.8\textsuperscript{11} & 2.40 \\
 & Oracle & 92.2\textsuperscript{3} & 89.1\textsuperscript{3} & 89.1\textsuperscript{3} & 4.34\textsuperscript{3} & 4.49\textsuperscript{1} & 84.4\textsuperscript{3} & 73.4\textsuperscript{6} & 4.24\textsuperscript{4} \\
 & Clean & 96.9\textsuperscript{1} & 93.8\textsuperscript{1} & 93.8\textsuperscript{1} & 4.48\textsuperscript{1} & 4.30 & 82.8\textsuperscript{1} & 76.6\textsuperscript{2} & 4.15\textsuperscript{1} \\
 & Noisy & 96.9\textsuperscript{1} & 92.2\textsuperscript{1} & 92.2\textsuperscript{1} & 4.29\textsuperscript{1} & 4.51\textsuperscript{1} & 87.5\textsuperscript{1} & 79.7\textsuperscript{2} & 4.20\textsuperscript{2} \\
\addlinespace[2pt]
DeepSeek & R0 & 100.0 & 96.9 & 96.9 & 4.81 & 1.94 & 12.5\textsuperscript{11} & 12.5\textsuperscript{11} & 2.53 \\
 & Oracle & 93.8\textsuperscript{3} & 92.2\textsuperscript{3} & 90.6\textsuperscript{3} & 4.36\textsuperscript{3} & 4.35\textsuperscript{2} & 75.0\textsuperscript{8} & 67.2\textsuperscript{10} & 4.20\textsuperscript{5} \\
 & Clean & 96.9\textsuperscript{1} & 95.3\textsuperscript{1} & 93.8\textsuperscript{1} & 4.62\textsuperscript{1} & 3.81 & 68.8\textsuperscript{4} & 64.1\textsuperscript{5} & 3.89\textsuperscript{1} \\
 & Noisy & 98.4 & 95.3 & 95.3 & 4.48 & 4.05\textsuperscript{1} & 73.4\textsuperscript{4} & 70.3\textsuperscript{4} & 4.01\textsuperscript{1} \\
\bottomrule
\end{tabular*}
\par\vspace{3pt}
\parbox{\textwidth}{\footnotesize
SF: source fidelity; HP: hard-constraint pass;
TA: task acceptance; TQ: task quality (1--5);
PA: preference alignment (1--5);
PR: preference realization; JS: joint success;
HM: mean of per-response harmonic scores (Trade-off only).
All rates are percentages. Superscripts indicate
unknown/missing judgments; rates are confirmed
events divided by 64. A dash denotes not applicable.}
\end{table*}


\begin{table*}[t]
\centering
\caption{Detailed Direct Conflict and Irrelevant results
(64 cases per model--profile cell).}
\label{tab:app_safety}
\setlength{\tabcolsep}{4.0pt}
\renewcommand{\arraystretch}{1.06}
\footnotesize
\begin{tabular*}{\textwidth}
{@{\extracolsep{\fill}}llrrrrrrr@{}}
\toprule
\multicolumn{9}{@{}l}{\textit{Direct Conflict}} \\
\addlinespace[2pt]
Model & Profile & SF & HP & TA & TQ & PA & PR & HOR \\
\midrule
GPT & R0 & 100.0 & 98.4 & 98.4 & 4.89 & 1.00 & 0.0 & -- \\
 & Oracle & 100.0 & 98.4 & 98.4 & 4.89 & 1.00 & 0.0 & 0.0 \\
 & Clean & 100.0 & 98.4 & 98.4 & 4.86 & 1.00 & 0.0 & 0.0 \\
 & Noisy & 96.9\textsuperscript{1} & 96.9\textsuperscript{1} & 95.3\textsuperscript{1} & 4.86\textsuperscript{1} & 1.11 & 1.6\textsuperscript{2} & 0.0\textsuperscript{3} \\
\addlinespace[2pt]
Gemini & R0 & 96.9 & 98.4 & 95.3 & 4.84 & 1.00 & 0.0 & -- \\
 & Oracle & 98.4 & 95.3 & 93.8 & 4.83 & 1.00\textsuperscript{1} & 0.0\textsuperscript{1} & 0.0\textsuperscript{1} \\
 & Clean & 98.4 & 87.5 & 87.5 & 4.69 & 1.06 & 1.6 & 1.6 \\
 & Noisy & 92.2\textsuperscript{3} & 84.4\textsuperscript{3} & 84.4\textsuperscript{3} & 4.67\textsuperscript{3} & 1.19 & 3.1\textsuperscript{2} & 1.6\textsuperscript{5} \\
\addlinespace[2pt]
DeepSeek & R0 & 100.0 & 96.9 & 96.9 & 4.81 & 1.00 & 0.0 & -- \\
 & Oracle & 96.9 & 85.9 & 85.9 & 4.44 & 1.48 & 10.9\textsuperscript{1} & 10.9\textsuperscript{1} \\
 & Clean & 93.8\textsuperscript{1} & 89.1\textsuperscript{1} & 89.1\textsuperscript{1} & 4.59\textsuperscript{1} & 1.31 & 7.8 & 7.8\textsuperscript{1} \\
 & Noisy & 96.9 & 85.9 & 85.9 & 4.44 & 1.64 & 14.1\textsuperscript{3} & 10.9\textsuperscript{3} \\
\addlinespace[3pt]
\midrule
\multicolumn{9}{@{}l}{\textit{Irrelevant}} \\
\addlinespace[2pt]
Model & Profile & SF & HP & TA & TQ & TPE & Leak & ILR \\
\midrule
GPT & R0 & 100.0 & 98.4 & 98.4 & 4.89 & 20.3\textsuperscript{5} & 0.0\textsuperscript{1} & -- \\
 & Oracle & 98.4 & 96.9 & 96.9 & 4.86 & 12.5\textsuperscript{16} & 0.0\textsuperscript{9} & 0.0\textsuperscript{10} \\
 & Clean & 95.3\textsuperscript{2} & 95.3\textsuperscript{2} & 95.3\textsuperscript{2} & 4.92\textsuperscript{2} & 17.2\textsuperscript{14} & 0.0\textsuperscript{5} & 0.0\textsuperscript{6} \\
 & Noisy & 96.9\textsuperscript{1} & 98.4\textsuperscript{1} & 96.9\textsuperscript{1} & 4.90\textsuperscript{1} & 21.9\textsuperscript{9} & 0.0\textsuperscript{3} & 0.0\textsuperscript{4} \\
\addlinespace[2pt]
Gemini & R0 & 96.9 & 98.4 & 95.3 & 4.84 & 18.8\textsuperscript{5} & 0.0\textsuperscript{1} & -- \\
 & Oracle & 100.0 & 98.4 & 98.4 & 4.86 & 10.9\textsuperscript{16} & 0.0\textsuperscript{9} & 0.0\textsuperscript{10} \\
 & Clean & 95.3\textsuperscript{2} & 89.1\textsuperscript{2} & 89.1\textsuperscript{2} & 4.74\textsuperscript{2} & 17.2\textsuperscript{14} & 0.0\textsuperscript{5} & 0.0\textsuperscript{6} \\
 & Noisy & 98.4\textsuperscript{1} & 95.3\textsuperscript{1} & 95.3\textsuperscript{1} & 4.87\textsuperscript{1} & 21.9\textsuperscript{9} & 0.0\textsuperscript{3} & 0.0\textsuperscript{4} \\
\addlinespace[2pt]
DeepSeek & R0 & 100.0 & 96.9 & 96.9 & 4.81 & 18.8\textsuperscript{5} & 0.0\textsuperscript{1} & -- \\
 & Oracle & 100.0 & 100.0 & 100.0 & 4.84 & 12.5\textsuperscript{16} & 0.0\textsuperscript{9} & 0.0\textsuperscript{10} \\
 & Clean & 96.9\textsuperscript{2} & 95.3\textsuperscript{2} & 95.3\textsuperscript{2} & 4.74\textsuperscript{2} & 17.2\textsuperscript{14} & 0.0\textsuperscript{5} & 0.0\textsuperscript{6} \\
 & Noisy & 98.4\textsuperscript{1} & 95.3\textsuperscript{1} & 95.3\textsuperscript{1} & 4.68\textsuperscript{1} & 21.9\textsuperscript{9} & 0.0\textsuperscript{3} & 0.0\textsuperscript{4} \\
\bottomrule
\end{tabular*}

\par\vspace{3pt}
\parbox{\textwidth}{\footnotesize
SF, HP, TA, and TQ are defined in
Table~\ref{tab:app_effectiveness}.
PR: preference realization;
HOR: harmful override;
TPE: target-preference expression (not necessarily leakage);
Leak: profile leakage;
ILR: induced leakage relative to R0.
TQ and PA are 1--5 scores; other values are percentages.
Superscripts count unknown or missing judgments.
HOR and ILR are undefined under R0 (shown as --).}
\end{table*}

\section{Success-Conditioned Quality Deterioration}
\label{app:evaluation-properties}

We compare recorded ordinal TQ grades on a common set of case--model pairs
with determined joint-success labels and grades under both conditions.
Let the ordered grades be $1,\ldots,K$, and write
$p_v(i)=\Pr(J_v=1,\mathrm{TQ}_v=i)$,
$F_v(k)=\sum_{i\leq k}p_v(i)$, and $a_v=F_v(K)$.
These are masses on the whole common cohort; they retain the proportion
of successes along with their grade distribution.
The report preserves $F_0,F_1$ and the paired fraction
$g=\Pr(J_0=0,J_1=1)$ entering success. The target is
\[
d=\Pr(J_0=J_1=1,\mathrm{TQ}_1<\mathrm{TQ}_0).
\]
We characterize its minimum over all joint grade tables compatible with
that report. The construction adapts one-sided stochastic-order
coupling~\citep{thorisson2000coupling,delbarrio2025invariant} to success-grade
masses, accounting explicitly for entry into the success set.

\subsection{Proof of Proposition~\ref{prop:profile-sensitivity}}

Let $t=a_1-g$ and let $\delta$ denote the right-hand side of
Equation~\ref{eq:profile-sensitivity-bounds}.

\paragraph{Lower bound.}
Each success at or below grade $k$ in view~1 is either a success at or
below $k$ in view~0, newly successful, or successful in both views with a
lower grade. Hence $F_1(k)\leq F_0(k)+g+d$ for every $k$, giving $d\geq\delta$.

\paragraph{Attainment.}
Compatibility gives $0\leq t\leq\min(a_0,a_1)$, and $F_1(k)\leq a_1$
implies $0\leq\delta\leq t$. Put $s=t-\delta$. Retain the lowest $s$ mass
of old successes and the highest $s$ mass of new successes, allocating
tied grades as needed. Their cumulative masses are
\[
\begin{aligned}
A(k)&=\min\{F_0(k),s\},\\
B(k)&=\max\{0,F_1(k)-g-\delta\}.
\end{aligned}
\]
Since $F_1(k)-F_0(k)\leq g+\delta$ and $B(k)\leq s$, we have $B(k)\leq A(k)$.
Pair these retained masses by increasing cumulative rank. An old grade
$i$ and a new grade $j<i$ at the same rank $u\in(0,s]$ would give
$A(j)<u\leq B(j)$, a contradiction. Thus this $s$ mass has no decline;
when $s=0$, skip this matching.

The remaining old and new success masses are $a_0-t+\delta$ and
$g+\delta$, each at least $\delta$. Choose and pair $\delta$ mass from
each, completing the stable-success mass $t$ with at most $\delta$
declines. Send the remaining old success mass $a_0-t$ to failure, receive
the remaining new success mass $g$ from failure, and put
$1-a_0-g\geq0$ in the failure--failure cell. This restores the reported
marginals and entry fraction. Therefore $d\leq\delta$; the lower bound
gives equality. For empirical counts, all selections and pairings can be
integer-valued.\hfill$\square$

\subsection{Information Preserved by Grouped Evaluation}

The result uses grade order rather than numerical distances between
grades. Any strictly increasing relabeling, with thresholds relabeled
accordingly, preserves the decline event and the bound.
A compatible table with no deterioration can still contain improvements;
the zero-decline condition concerns the direction specified by $d$.

For a fixed group $G$ shared by both ends of each pair, let $w_r=\Pr(G=r)$,
and define $F_{v,r}(k)$ and $g_r$ on the corresponding conditional cohort.
Retaining these grouped reports gives the sharp minimum
\begin{equation}
\begin{split}
d_{\min}^{\mathrm{group}}
&=\sum_r w_r\max\bigl\{0,\\
&\quad\max_k[F_{1,r}(k)-F_{0,r}(k)]-g_r\bigr\}.
\end{split}
\label{eq:group-quality-changes}
\end{equation}
It is at least the pooled minimum, by convexity of the maximum of these
linear expressions. Applying the endpoint construction within each group
and combining it with the same weights establishes sharpness.
Thus retaining relation or model groups can prevent opposing distributional
movements from canceling in the report.

\subsection{Empirical Specialization}

For Compatible and Tradeoff Clean--Noisy, 345 of 384 pairs have determined
$J$ and TQ at both ends. The other 39 retain their unresolved judgments.
On this common cohort, $g=21/345$ and the maximum cumulative difference is
$23/345$ at grade~4. The report therefore forces at least two stable-success
declines. Full grade matching identifies 41 declines among 268 stable
successes, giving the conditional rate $41/268=15.3\%$; 25 pairs improve
and 202 retain their grade. The proposition's $d$ uses all 345 pairs as
its denominator, while this conditional rate uses the 268 stable successes.

For Tradeoff alone, 170 pairs have both judgments determined. Each view
contains 127 successes, but 14 pairs enter success and 14 leave.
The cumulative grade mass at or below 4 increases by $19/170$,
exceeding the entrant mass $14/170$. At least five stable-success pairs
must therefore decline. Full matching identifies 27 declines among 113
stable successes. This contrast separates an unchanged success rate,
replacement within the success set, and deterioration among persistent
successes.

For the separate realization contrast in
Section~\ref{sec:model-differences}, 173 of 192 Tradeoff pairs have
determined realization labels. Counts $(25,11,11,126)$ give equal gains
and losses, with 22 status changes. Its cohort and diagnostic event are
distinct from the TQ analysis above.

\end{document}